\RequirePackage{fix-cm}
\documentclass[twocolumn,epjc3]{svjour3}

\RequirePackage{color,graphicx,url}

\journalname{Eur. Phys. J. A}

\usepackage{amsfonts,amsmath,amssymb,bm,xspace}
\usepackage{authblk}
\graphicspath{{./figures/}}

\usepackage[colorlinks=true]{hyperref}  

\graphicspath{{figures/}} 
\usepackage{multirow}

\usepackage{academicons} 
\newcommand{\orcid}[1]{\href{https://orcid.org/#1}{\textcolor[HTML]{A6CE39}{\aiOrcid}}}
\usepackage{xcolor}

\usepackage{bm}
\usepackage{enumitem}

\usepackage{nccmath}

\usepackage{xfrac}

\hypersetup{%
    pdfsubject=Paper,
    pdfkeywords={nuclear physics} {MBPT} {chiral EFT} {asymmetric nuclear matter}
    unicode=true,
    breaklinks=true,
     colorlinks   = true,
     linkcolor = blue,
     citecolor = blue,
     menucolor = blue,
     urlcolor = blue
}

\begin{document}

\title{Scalar field, nucleon structure and relativistic chiral theory for nuclear matter}
\dedication{
Dedicated to the memory of Peter Schuck}

\author{Guy Chanfray}
\institute{Univ Claude Bernard Lyon 1, CNRS/IN2P3, IP2I Lyon, UMR 5822, F-69622, Villeurbanne, France }




\date{Draft version:\today, Received: date / Accepted: date}

\maketitle

%
\abstract{The work that Peter Schuck and I carried out during the nineties in collaboration with the Lyon and Darmstadt theory groups is summarized. I retrace how our theoretical developments combined with experimental results concerning  the in-medium modification of the pion-pion interaction  allowed a clarification of  the chiral status of the sigma meson introduced in relativistic theories of nuclear matter. This enabled us to build  a relativistic chiral theory, now called  the chiral confining model  of nuclear matter, which includes the effect of the nucleon  substructure, namely the response of the nucleon to the nuclear scalar field  generating an efficient and natural contribution to the saturation mechanism. Using parameters from a QCD-connected version of the chiral confining model, I  describe the relative roles of the chiral  scalar field and  two-pion (or two-rho) exchange for the in-medium $NN$ attractive interaction and the associated sources  of three-body interactions needed for the saturation mechanism.}
\PACS{
{Chiral Lagrangians}
\and
{confinement}
\and
{Nuclear matter}
} 

\section{Introduction}
\label{sec:intro}
In the early nineties, Peter Schuck and I  published with  other collaborators from the Lyon and Darmstadt groups a set of papers concerning the in-medium modification of the $\pi\pi$ interaction and of the associated resonances, namely 
the rho meson and this very elusive object known as the  sigma meson \cite{Schuck1988,Chanfray1990,Chanfray1991,Chanfray1992c,Chanfray1993,Aouissat1993,Aouissat1995,Rapp1999,Aouissat2000}. In both cases a quite spectacular accumulation of strength was obtained
in the associated spectral functions at low energy, just above the two-pion threshold. Concerning the rho meson we  predicted a rather dramatic change in its spectral function  when the density is increased \cite{Chanfray1992c,Chanfray1993}. A few years later I, in collaboration with R. Rapp and J. Wambach, incorporated this effect in a dynamical approach to explain the dilepton production data in relativistic heavy ion collisions of the CERES collaboration at CERN/SPS \cite{Chanfray1996,Rapp1998}. This scenario, once completed by the direct coupling of the rho to some N*h excitations \cite{Urban,RWH00}, allowed to give an interpretation to the broadening of the dilepton spectrum observed by the NA60 collaboration \cite{NA6006}. We also showed that this broadening that we have predicted could be interpreted as associated with a mechanism of chiral symmetry restoration through a mixing of vector and axial correlators  \cite{Chanfray1998,Rosaclot,RWH00,OSAKA,PANIC}.\\

Although these works related to the in-medium rho meson had a notable repercussion, I will rather discuss in what follows the properties of the sigma meson and consequently the question of the role of the scalar field in nuclear physics. The in-medium scalar-isoscalar modes in nuclei have been studied experimentally through two-pion production in reactions induced either by pions \cite{Bonnuti,Staros} or photons \cite{TAPS} on various
nuclei. All these collaborations have observed a systematic A dependent downwards shift of the strength when the pion pair is produced in the sigma meson channel. These results brought considerable attention in the nuclear physics community.  Then came three related questions:
\begin{enumerate}
    \item Do these spectacular medium effects affect the exchange of the  “sigma meson” in nuclear matter? 
    \item What is the relationship with the “nuclear physics sigma meson” of relativistic mean field theory? 
    \item What is the real nature of this "nuclear physics sigma meson", regarding the constraints from chiral symmetry?
\end{enumerate}
To answer the second an third questions let us consider the relativistic mean-field approaches initiated by Walecka and collaborators \cite{SerotWalecka1986,Walecka1997} where the nucleons move in an attractive scalar and a repulsive vector background fields.
It provides an economical saturation mechanism and a well-known success is the correct magnitude of the spin-orbit potential where the large vector and scalar fields contribute in an additive way. Now the question of the very nature of these background fields has to be elucidated. It is highly desirable to clarify their relationship with the QCD condensates, in particular the chiral quark condensate, and more generally with the low energy realization of chiral symmetry which is spontaneously broken in the QCD vacuum and is expected to be progressively restored when the density increases. If the origin of the repulsive vector field can be safely identified as associated with the omega vector-meson exchange, the real nature of the attractive Lorentz scalar field has been a controversial subject since there is no sharp scalar resonance with a mass of about $500-700$ MeV, which would lead to a simple interaction based on a scalar particle exchange.\\
To bridge the gap between relativistic theories of the  Walecka type and approaches insisting on chiral symmetry, one has to map the "nuclear physics sigma meson" of the Walecka model (let us call it $\sigma_W$ from now on) with a chiral quantity. For instance, one may be tempted to identify $\sigma_W$ with $\sigma$, the  chiral partner of the pion. It is however forbidden by chiral constraints. This point has been first addressed by Birse \cite{Birse}: it would lead to the presence of terms of order $M_\pi$ in the $NN$ interaction which is not allowed. The other possibility is to formulate an  effective theory parametrized in terms of the field associated with the fluctuations of the chiral condensate in a $SU(2)$ matrix form, $W=\sigma + i\vec{\tau}\cdot\vec{\pi}$,
by going from cartesian to polar coordinates, i.e., going from a linear to a non linear representation : $W=\sigma + i\vec{\tau}\cdot\vec{\pi}\equiv S\, U\equiv (s\, +\, F_{\pi})\,U\equiv (\sigma_W +\, F_{\pi})\,e^{i\,{\vec{\tau}\cdot\vec{\phi}(x)}/{F_\pi}}$. It was instead   proposed and justified  in Ref.~\cite{Chanfray2001} to identify $\sigma_W$, not with $\sigma$, but  with the chiral invariant $s=S-F_\pi$ field associated with the radial fluctuation of the chiral condensate, $S$, around the "chiral radius" $F_\pi$, identified with the pion decay constant.  The sigma and the pion, associated with the amplitude  $s\equiv\sigma_W$ and phase fluctuations $\vec{\phi}$ of this condensate, are considered in our approach to be effective degrees of freedom. Their dynamics are governed by an effective chiral potential, $V_\chi\left(\sigma,\vec{\pi}\right)$, having a typical Mexican hat shape associated with a broken (chiral) symmetry of the QCD vacuum. This proposal, which gives a plausible answer to the long standing problem of the chiral status of Walecka theories, has also the merit of respecting all the desired chiral
constraints. In particular the correspondence $s\equiv \sigma_W$ generates a coupling of the scalar field to the derivatives of the pion field. Hence the radial mode decouples from low-energy pions (as the pion is a quasi-Goldstone boson) whose dynamics is governed by chiral perturbation theory. It follows that the answer to the first question is negative. The strong medium effect seen in $2\pi$ production experiment is associated with the dressing of the $\sigma$ propagator by in-medium modified $2\pi$ states, the single pion line being replaced by a pion branch, a collective mixture of pions, p-hole and $\Delta$-hole states. However, this in-medium modified sigma exchange between nucleon has to be completed by other exchanges with resulting delicate compensations. Their origin is the well-known pair suppression, in the case of pseudo-scalar coupling, by $\sigma$ exchange for the $\pi N$ amplitude. This translates into the elimination of the sigma dressing in the $NN$ interaction. We have explicitly checked that this cancellation holds to all orders in the dressing of the sigma. The net result amounts to the exchange of the $s$ field which is essentially free of medium modification since it decouples from the pion field $\vec{\phi}$ which itself has  a pseudovector coupling to the nucleon. This is this chiral invariant object which is identified with the "nuclear physics sigma meson". A detailed discussion of this somewhat subtle topic is given in \cite{Chanfray2001,Martini2006} but it can be concisely summarized by the figure 3 of \cite{Martini2006}. Due to the aforementioned compensations the total "correlated $2\pi$--pion chiral partner $\sigma$" reduces to the $s\equiv \sigma_W$ exchange if we neglect the tiny contribution from $\pi\pi$
interaction which vanishes in the chiral limit.
\section{Chiral symmetry and nucleon structure}
In the early 2000s, we  realized  that the above construction faces with two major problems related to nuclear stability and the underlying nucleon structure.\\
Concerning the first point there is a well identified problem regarding the nuclear saturation with usual chiral effective theories \cite{Boguta83,KM74,BT01,PANIC}: independently of the particular chiral model, in the nuclear medium the value of the scalar field $S$ ($\equiv S_{medium}$) is different from the one in vacuum ($\equiv S_{vacuum}$, the minimum of  the vacuum effective potential represented by a "Mexican hat" potential). At $S_{medium}$ the chiral potential has a smaller curvature : $V''(S_{medium}) < V''(S_{vacuum})$. This single effect results in the lowering of the sigma mass and destroys the stability, which is a problem for the applicability of such effective theories in the nuclear context. The effect can be associated with a $s^3$ tadpole diagram  generating attractive three-body forces destroying saturation even if the repulsive Z graph from the Walecka mechanism is present. \\
The second point is associated with the chiral properties of the nucleon, namely the pion-nucleon sigma term and the chiral susceptibility of the nucleon. According to the lattice data  analysis of the Adelaide group, \cite{LTY03,LTY04,TGLY04,AALTY10}, the nucleon mass can be expanded in terms of the pion mass squared, $M^2_{\pi}$, as $M_N(M^2_{\pi}) = 
a_{0}\,+\,a_{2}\,M^2_{\pi}\, +\,a_{4}\,M^4_{\pi}\,+ ...+\,\Sigma_{\pi}(M^2_{\pi},\, \Lambda)$ 
 where the pionic self-energy, $\Sigma_{\pi}(M^2_{\pi}, \Lambda)$, containing the non analytical contribution, is explicitly separated out. The latter is calculated with just  one adjustable cutoff parameter $\Lambda$ entering the $\pi NN, \pi N\Delta$ form factor regularizing the pion loops. While the $a_2$ parameter is related to the non pionic piece of the $ \pi N$ sigma term, $a_4$ is related to the nucleon QCD scalar susceptibility. The important point is that $a_4\simeq -0.5\,GeV^{-3}$ is essentially compatible with zero in the sense that it is
much smaller than in a chiral effective model, $(a_4)_{L\sigma M} =-F_\pi\,g_{S}/2 M^4_{\sigma }\simeq -3.5\,GeV^{-3}$, where the nucleon is seen as a juxtaposition of three constituent quarks getting their mass from the chiral condensate \cite{Ericson2007} ($g_S$ is the scalar coupling constant of the nucleon and $M_\sigma$ is the sigma mass in the linear sigma model). \\
The common origin of these two failures can be attributed to the absence of confinement. In reality the composite nucleon responds to the nuclear environment, i.e., by readjusting its confined quark structure as pointed out in the pioneering paper of P. Guichon \cite{Guichon1988} at the origin of the Quark Meson Coupling (QMC) model \cite{Guichon2004}. The resulting polarization of the nucleon can be accounted for by the phenomenological introduction of the positive scalar nucleon response in the nucleon mass evolution. The physical motivation to introduce this nucleonic response is the observation that nucleons experience huge fields at finite density, e.g., the scalar field is of the order of a few hundred of MeV at saturation density. Nucleons, being in reality composite objects, will react against the nuclear environment (i.e., the background nuclear scalar field) through a (self-consistent) modification of the quarks wave functions.

\section{The  chiral confining model}
\begin{figure}
		\centering
		\includegraphics[width=0.5\textwidth]{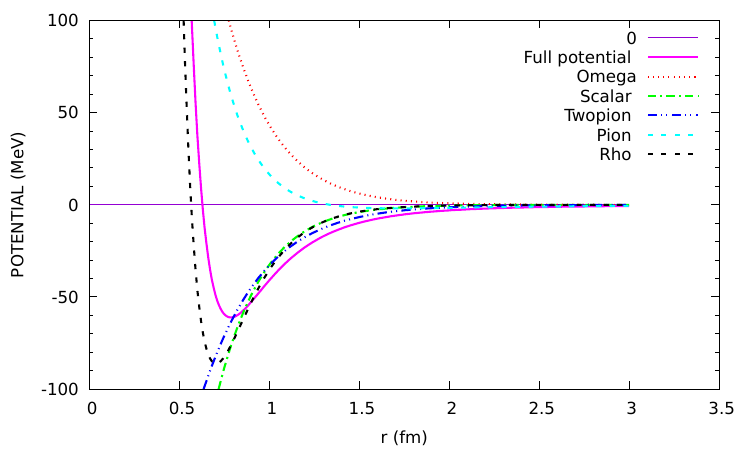}
		\caption{ The full bare  NN  potential  (full line) and the various contributions. The value of the parameters are those given in the text of section 4.}
		\label{POTNN}
	\end{figure}

These considerations led the development of a phenome-nological model \cite{Chanfray2005,Chanfray2007,Massot2008,Massot2009,Rahul}, that we now call the chiral confining model, where we complemented the relativistic chiral approach in such a way that the effect of the nucleon response is able to counterbalance the attractive chiral tadpole diagram to get good saturation properties, especially the correct curvature coefficient - the incompressibility empirical parameter, $\kappa_{sat}$. It is described by a lagrangian, with standard Yukawa couplings of the nucleon to various mesonic fields,
$
{\cal L}=\bar\Psi\,i\gamma^\mu\partial_\mu\Psi\,+\,
{\cal L}_s\,+\,{\cal L}_\omega\,+\,{\cal L}_\rho\,\,+\,{\cal L}_\pi
$,
with:
\begin{eqnarray}
{\cal L}_s &=& -M^*_N(s)\bar\Psi\Psi\,-\,V(s)\,+\,\frac{1}{2} \partial^\mu s\,\partial_\mu s\nonumber\\	
{\cal L}_\omega &=& - g_V\,\omega_\mu\,\bar\Psi\gamma^\mu\Psi\,+\,\frac{1}{2}\,m^2_V\,\omega^\mu\omega_\mu
\,-\,\frac{1}{4} \,F^{\mu\nu}F_{\mu\nu}\nonumber\\	
{\cal L}_\rho &=&- g_\rho\,\rho_{a\mu}\,\bar\Psi\gamma^\mu \tau_a\Psi
\,-\,g_\rho\frac{\kappa_\rho}{2\,M_N}\,\partial_\nu \rho_{a\mu}\,\Psi\bar\sigma^{\mu\nu}_{}\tau_a\Psi\nonumber\\
&&\,+\,\frac{1}{2}\,m^2_\rho\,\rho_{a\mu}\rho^{\mu}_{a}
\,-\,\frac{1}{4} \,G_a^{\mu\nu}G_{a\mu\nu}\nonumber\\	
{\cal L}_\pi &=& \frac{g_A}{2\,F_\pi}\,\partial_\mu\phi_{a\pi}\bar\Psi\gamma^\mu\gamma^5\tau_a\Psi-\,\frac{1}{2}\,m^2_{\pi}\phi_{a\pi}^2
\nonumber  \label{LAG2}\\
&&\,+\,\frac{1}{2}\, \partial^\mu\phi_{a\pi}\partial_\mu \phi_{a\pi}.
\end{eqnarray}
It involves  the scalar field $s$ (the "nuclear physics sigma meson" $\sigma_W$), the pion field $\phi_{a\pi}$, and the vector fields associated with the omega meson  channel ($\omega^\mu$) and with  the rho meson channel ($\rho_a^\mu$). Each meson-nucleon vertex is regularized by a monopole meson-nucleon form factors (with cutoff $\Lambda_{S,\pi,V,\rho}$) mainly originating from the compositeness of nucleon hence generating a bare OBE Bonn-like $NN$ interaction. This potential, completed by a fictitious $\sigma'$ meson exchange simulating the effect of two-pion exchange with $\Delta$'s in the intermediate state, is represented in fig. \ref{POTNN}.
 This is a rather standard lagrangian but with no ad hoc density dependent coupling constant. Instead, the three (multi)-body forces are generated by two specific crucial ingredients beyond the simplest approach:  the introduction of an  in-medium modified nucleon mass which is supposed to embed its quark substructure  and the  presence of a chiral effective potential associated with the chirally broken vacuum. 
The effective Dirac nucleon mass $M^*_N(s)$ deviates from the bare nucleon mass in presence of the nuclear scalar field $s$:
\begin{eqnarray}
		M^*_N(s)&=& M_N + g_S\, s +   \frac{1}{2}\kappa_{NS} \,s^2 + \mathcal{O}(s^3). 
  \end{eqnarray}
In \cite{Chanfray2005,Martini2006,Chanfray2007,Massot2008,Massot2009,Massot2012}, we took the pure linear sigma model value ($g_S=M_N/F_\pi$) for the scalar coupling constant but in the recent work \cite{Rahul}, $g_S$ was allowed to deviate from the L$\sigma$M and was fixed by a Bayesian analysis. This quantity actually corresponds to the first order response of the nucleon to an external scalar field and can be obtained in an underlying microscopic model of the nucleon. The nucleon scalar  susceptibility $\kappa_\mathrm{NS}$ is another response parameter which reflects the polarization of the nucleon, i.e., the self-consistent readjustment of the quark wave function in presence of the scalar field.    Very generally the scalar coupling constant, $g_S$, and the nucleon response parameter,  $\kappa_{NS}$, depend on the subquark structure and the confinement mechanism as well as the effect of spontaneous chiral symmetry breaking.
In our previous works \cite{Chanfray2005,Chanfray2007,Massot2008,Massot2009,Rahul,Massot2012} we introduced a dimensionless parameter, 
\begin{equation}
 C\equiv \frac{\kappa_\mathrm{NS}\,F_\pi^2}{2 M_N},   
\end{equation}
which is expected to be of the order $C\sim 0.5$ as in the MIT bag  used in the QMC framework. Such a positive response parameter can be generated  if confinement dominates spontaneous chiral symmetry breaking in the nucleon mass origin, as discussed in Ref. \cite{Chanfray2011} within particular models.\\
In the majority of our previous works \cite{Martini2006,Chanfray2005,Chanfray2007,Massot2008,Massot2009,Rahul,Massot2012}, the chiral effective potential had the simplest linear sigma model (L$\sigma$M) form:
 \begin{equation}
V_{\chi,{L\sigma M}}(s)=\frac{1}{2}\,M^2_\sigma \,s^2\, +\,\frac{1}{2}\frac{M^2_\sigma -M^2_\pi}{ F_\pi}\, s^3\,+\,
\frac{1}{8}\,\frac{M^2_\sigma -M^2_\pi}{ F^2_\pi} \,s^4 \label{eq:VLSM}.
\end{equation} 
 Combining the effects of the tadpole diagram  (i.e, assocciated with the $s^3$ in the chiral potential) and of the response parameters, it is possible to show that the scalar sector generates a three-nucleon contribution to the energy per nucleon:
\begin{eqnarray}
E^{(3b-s)}&\simeq&\frac{g^2_S}{2\,M^4_\sigma}\,\left(\kappa_\mathrm{NS}-\frac{g_S}{F_\pi}\right)\,\rho^2_s\,\nonumber\\
&=&\,\frac{g^3_S}{2\,M^4_\sigma\,F_\pi}\,\left(2\,\frac{M_N}{g_S\,F_\pi}\,C\,-\,1\right)\,\rho^2_s .
\label{eq:THREEBOD}
\end{eqnarray}
This is the result already quoted in Eq.~(44) of Ref.~\cite{Ericson2007}, but without the factor $M_N/g_S F_\pi$ which appears when the scalar coupling constant is allowed to deviate from its pure L$\sigma$M value as in Ref. \cite{Rahul}. If the confinement ($C$)  dominates over the  attractive chiral attraction (tadpole diagram), it  provides a very natural saturation mechanism but at variance with the QMC model which ignores the attractive tadpole diagram present in the chiral approach, this requires à $C$ parameter close to or even larger than one \cite{Chanfray2005,Chanfray2007,Massot2008,Massot2009,Rahul,Massot2012,Cham}. \\
The explicit connection between lattice QCD  parameters $a_2, a_4$ and the response parameter $g_S, C$, in the L$\sigma$M case has been given in  \cite{Chanfray2007,Massot2008}: 
\begin{equation}
a_2= \frac{F_\pi\, g_{S}}{M^2_{\sigma}}, 
\qquad
a_4 =-\frac{F_\pi\,g_{S}}{2 M^4_{\sigma }}\,\left(2\,\frac{M_N}{g_S\,F_\pi}\,C\,-\, 3\right).\label{LATTRES}
\end{equation}
Notice that in the expression of $a_4$  the factor $M_N/F_\pi g_S$ in front of $C$, present in our recent paper \cite{Rahul}, was absent in \cite{Chanfray2007,Massot2008} since the nucleon mass was fixed to be $M_N=F_\pi g_S$. Depending of the details of the chiral extrapolation, the extracted values of the parameters are within the range in between  $a_2\simeq 1.5 \,GeV^{-1},\,a_4\simeq -0.5 \,GeV^{-3}$ \cite{LTY04} and $a_2\simeq 1.\,GeV^{-1},\,a_4\simeq -0.25\, GeV^{-3}$ \cite{LTY03,AALTY10}. The quantity $a_2 M^2_\pi \sim 20-30\,MeV$ represents the non pionic piece of the sigma commutator directly associated with the scalar field, $s$ (see the detailed discussion of this quantity in Ref. \cite{Chanfray2007}).
One very robust conclusion   is that the lattice result for $a_4$ is much smaller than the one obtained in  the simplest  linear sigma model ignoring the nucleonic response ($C=0$), for which $a_4\simeq -3.5\,GeV^{-3}$. Hence  Lattice data require a strong compensation from effects governing the three-body repulsive force needed for the saturation mechanism: compare Eqs. \eqref{LATTRES} and \eqref{eq:THREEBOD}.
Consequently both lattice data constraints and nuclear matter phenomenology require a quite large value of the dimensionless response parameter, $C$, at least larger than one. Moreover in a recent work based on a Bayesian analysis with lattice data as an input \cite{Rahul}, we found that the response parameter is strongly constrained to a value $C\sim 1.4$ very close to the value where the scalar susceptibilities changes of sign: $C=1.5$.\\
This model has been  applied in the past to the equation of state of nuclear matter \cite{Chanfray2005,Chanfray2007,Massot2008,Massot2009,Rahul,Cham} and neutron stars \cite{Massot2008,Massot2012,Cham} as well as to the study of chiral properties of nuclear matter \cite{Chanfray2005,Martini2006,Ericson2007,Chanfray2007} at different levels of approximation in the treatment of the many-body problem (RMF, Relativistic Hartree Fock or RHF, pion loop correlation energy). Among other results, one important lesson is the importance of a coherent treatment of the Fock term including the rearrangement terms \cite{Massot2008,Cham}  which has to be consistently incorporated at the level of the self-energies and total energy. Although this problem is of minor importance for what concerns the binding energy around nuclear matter density, the use of a Hartree-Fock basis in place of the Hartree basis for the nucleon Dirac wave function, may play a very important role at high density in limiting the maximum mass of hyperonic neutron star as pointed out in Ref. \cite{Massot2012}.\\

The problem one has to face is that it seems impossible to find a realistic confining model for the nucleon able to generate $C$ larger than one required by lattice data and saturation properties of nuclear matter. As pointed out in our recent paper \cite{Chanfray2023}, one possible reason for this discrepancy between models and phenomenological values of $C$ lies in the  use of the linear sigma model (L$\sigma$M) which  is probably too naive. Hence one should certainly use a enriched chiral effective potential from a model able to give a correct description of the low energy realization of chiral symmetry in the hadronic world.
A good easily tractable candidate is the Nambu-Jona-Lasinio (NJL) model defined by the Lagrangian given in Eq. (9) of \cite{Chanfray2023} or Eq. (59) of \cite{Universe}: it depends on four parameters: the coupling constants $G_1$ (scalar), $G_2$ (vector), the current quark mass $m$ and a (non covariant) cutoff parameter $\Lambda$. Three of these parameters ($G_1$, $m$, and $\Lambda$) are adjusted to reproduce the pion mass, the pion decay constant and the quark condensate, ignoring pion-axial mixing.  We refer the reader to \cite{Chanfray2011,Chanfray2023} for more details.
As established in \cite{Chanfray2023}, the net effect of the use of the NJL model is to replace the L$\sigma$M chiral potential by its NJL equivalent which very well approximated by its expansion to third order in $s$:
\begin{equation}
V_{\chi,\mathrm{NJL}}(s)= \frac{1}{2}\,M^2_\sigma\, {s}^2\, +\,\frac{1}{2}\,\frac{M^2_\sigma -M^2_\pi}{ F_\pi}\, {s}^3\,\big(1\,-\,C_{\chi,\mathrm{NJL}}\big) +...\, .\label{eq:vchiNJL}
\end{equation}
The effective sigma mass $M_\sigma \sim 2 M_0$ ($M_0$ being the vacuum value of the constituent quark mass) as well as the pion mass $M_\pi$ and $F_\pi$ are calculated within the model.  This constitutes à new version of the model, called the NJL chiral confining model, as proposed in \cite{Chanfray2023} and in another recent paper \cite{Universe}. The main difference with the original phenomenological version using the L$\sigma$M  lies in the presence 
of the $C_{\chi,\mathrm{NJL}}$ parameter whose expression in terms of a NJL loop integral is given in \cite{Chanfray2023}. For typical values of the  NJL parameters its value is in the range $C_{\chi,\mathrm{NJL}}: 0.4-0.5$.  The effect of the parameter $C_{\chi,\mathrm{NJL}}$ is thus to reduce the attractive tadpole diagram and make the chiral potential more repulsive. The net visible effect of this enriched chiral potential is to modify the three-body force contribution to the the energy per nucleon generated by the scalar sector initially given by Eq. \eqref{eq:THREEBOD}, according to: 
\begin{equation}
E^{(3b-s)}
=\frac{g^3_S}{2\,M^4_\sigma\,F_\pi}\,\left(2\, \left[\frac{M_N}{g_S\,F_\pi}\,C +\,\frac{1}{2}\,C_\chi\right]\,-\,1\right)\,\rho^2_s.
\label{eq:THREEBODN}
\end{equation}
The lattice QCD parameter $a_4$ is as well modified according to (compare with Eq. \eqref{LATTRES}):
\begin{equation}
a_4 =\frac{F_\pi\,g_{S}}{2 M^4_{\sigma }}\,\left(2\,\left[\frac{M_N}{g_S\,F_\pi}\,C +\,\frac{3}{2}\,C_\chi\right]\,-\,3 \right).\label{eq:LATTIX}
\end{equation}
The important conclusion discussed in detail in Ref. \cite{Chanfray2023} is that for a given three-nucleon force allowing saturation one needs a lower value of the dimensionless parameter $C$ to obtain a small value of  the lattice parameter $a_4$ compatible with lattice data.
Said slightly differently, lattice data (small value of $a_4$ ) become compatible with models value for $C$ while preserving  the three-body repulsive forces needed for the saturation mechanism.\\

 A plausible physical picture underlying our approach can be summarized as follows: nucleons are viewed as Y shaped-strings with constituent quarks at the ends moving in a in-medium modified broken vacuum.  Hence, what is usually called the "nuclear medium" can be seen as a "shifted vacuum" with lower value  of the constituent quark mass which coincides  with the in-medium expectation value, $M=\bar{\mathcal{S}}(\rho)$, of the chiral invariant scalar field, $\mathcal{S}=(M_0/F_\pi)\,(s+F_\pi)$, associated with the radial fluctuation mode of the chiral condensate. This idea has been implemented in our recent paper \cite{Universe}  in a framework  inspired from the field correlator method (FCM) developed by Y. Simonov and collaborators \cite{Simonov1997,Tjon2000,Simonov2002,Simonov2002a,Simonov-light}.
By performing a gluon averaging in the euclidean QCD partition function, on can generate an effective non perturbative interaction between quarks mediated by a gluon correlator parametri-zed in a convenient gaussian form:
\cite{Tjon2000,Simonov2002}: 
\begin{equation}
D(x)==\frac{\sigma}{2\pi T^2_g} \, e^{-x^2/4 T^2_g}\,\,\,\,\hbox{with}\,\,\,\,\,T_g=\sqrt{\frac{9\sigma}{\pi^3 \mathcal{G}_2}}.
\end{equation}
It depends on two  QCD quantities measured in lattice QCD \cite{Digiacomo}, namely the string tension $\sigma=0.18$~GeV$^2$ and the gluon correlation length, $T_g=0.25\div 0.3\, fm$ itself related to the gluon condensate $\mathcal{G}_2$. In presence of a string junction (or a static heavy quark) placed in some point $\bf{r}_0$ in the QCD vacuum, multiple gluon correlations will generate  a string with width $T_g$ which develops  between the point $\bf{r}_0$ and the two light quarks located in ${\bf x}$ and ${\bf y}$ with ${\bf r}={\bf x}-{\bf y}$ and and $\bf{R}=(\bf{x}+\bf{y})/2\,-\,\bf{r}_0$, the string length variable. As explained in detail 
in section 3.3 of Ref. \cite{Universe}, modulo some ansatz prescription, this approach allows  to generate simultaneously, at a semi-quantitative level,  a confining interaction built from the gluon correlator with long distance ($R\gg T_g$)  behaviour $V_C(R)=\sigma \, R$,    together with an equivalent NJL model  with scalar interaction strength $G_1=120\pi\sigma T^4_g/(4N_cN_F)\sim 10$~GeV$^{-2}$ and cutoff $\Lambda\sim 1/T_g\sim 600$~MeV. 
If we take  accepted values of the string tension, $\sigma=0.18\,\mathrm{GeV}^2$, and gluon condensate,  $\mathcal{G}_2=0.025\,\mathrm{GeV}^4$, which implies a gluon correlation length, $T_g= \sqrt{9\sigma/\pi^3\mathcal{G}_2}=0.286\,\mathrm{fm}$, one obtains $G_1=12.514 \,\mathrm{GeV}^{-2}$.  We thus fix $\Lambda =0.604\, MeV$ and a current quark mass, $m=5.8\,MeV$, to obtain $M_0=356.7\,MeV$, $F_\pi=91.9\,MeV$, $M_\pi=140\,MeV$,   $\langle \Bar{q} q\rangle=-\left(241.1\,MeV\right)^3$. It is remarkable that taking accepted values of the two basic QCD parameters, namely the string tension and the gluon condensate, one recovers the NJL parameters yielding the good NJL phenomenology, or at least their order of magnitude. We also find that the NJL parameter entering the cubic term of the chiral effective potential is significant, $C_\chi =0.488$, and the effective sigma mass parameter is $M_\sigma=716.4\,MeV$.\\

It is also  possible to establish a bound state equation  \cite{Universe} which is equivalent to a Dirac equation for a constituent quark with mass $M=\bar{\mathcal{S}}(\rho)$ moving in a BCS-like state representing the (modified) broken  QCD vacuum and submitted to the confining potential with origin at the string junction point (see also \cite{BBRV98}). The nucleon wave function is just the product of the three quark orbitals properly projected on to the color singlet state with $I=J=1/2$. After center of mass correction the quark core contribution to the bare nucleon mass is found to be $M_N^{(core)}=3.62\,\sqrt{\sigma}$. The response parameters are finally obtained as:
\begin{eqnarray}
g_S &=&\frac{M_0}{F_\pi}\left(\frac{\partial M_N^{(core)}(\mathcal{S})}{\partial\mathcal{S} }\right)_{\mathcal{S}=M_0}=6.52,\\
C&=&\frac{M_0^2}{2 M_N}\left(\frac{\partial^2 M_N^{(core)}(\mathcal{S})}{\partial\mathcal{S}^2 }\right)_{\mathcal{S}=M_0}= 0.32.
\end{eqnarray}
The parameters entering the three-body force (Eq.\eqref{eq:THREEBODN}) and the chiral susceptibility (Eq. \eqref{eq:LATTIX}) are: $$\Tilde{C}_3=\frac{M_N}{g_S F_\pi}\,C +\frac{1}{2}\,C_\chi=0.75$$ and 
 $$\Tilde{C}_L=\frac{M_N}{g_S F_\pi}\,C +\frac{3}{2}\,C_\chi=1.23.$$
For the comparison with lattice data our model calculation yields: $a_2=1.175\,GeV^{-1},\,   a_4=-0.615\,GeV^{-3}$.
The value of $a_2$ is compatible with lattice data and yields a non pionic contribution to the sigma commutator $\sigma^{(s)}=a_2\,M^2_\pi=23.5\,MeV$. 
The value of $a_4$, although a bit high, is essentially compatible with lattice data in the sense that it is much smaller than the value obtained in  the simplest  linear sigma model ignoring the nucleonic response (i.e., $C=0$) and the NJL correction (i.e., $C_\chi=0$), for which $a_4\simeq -3.5\,GeV^{-3}$. Hence the model generates the strong compensation required by lattice data from effects governing the three-body repulsive force needed for the saturation mechanism: compare
Eqs. \eqref{eq:LATTIX} and \eqref{eq:THREEBODN}. It is also possible to calculate the core rms and the vertex form factors in the various (scalar, axial and vector) Yukawa channels. The cutoff values for the equivalent monopole form factors regularizing the corresponding Yukawa couplings to the nucleon  are all close to  $\Lambda =1\,GeV$. 
\section{The relative roles of the scalar field and two -pion exchange in the nuclear matter equation of state}
Already in the early nineties, Peter Schuck, Wolfgang N\"{o}renberg  and I started  to think about the effect of the in-medium modified two-pion  (and  two-rho) exchange  on the nuclear equation of state. In Ref. \cite{CSN1990}  we used a model hamiltonian containing pion and rho  exchanges completed by a quartic interaction in the $NN$ spin-isospin channel  characterized by a Landau-Migdal $g'_{NN}$ parameter which was also extended to the $N\Delta$ an $\Delta\Delta$ interactions. This effect of the pion and rho loops in presence of short-range correlations on top of the mean field was calculated using the text-book charging formula but generalized to a non static (pion and rho exchanges) interactions. Such a calculation contains not only  the effect of RPA-like long-range correlations but also includes short-range correlations through the aforementioned screening of the pion and rho exchanges.  Using Green's function  techniques \cite{CSN1990} and with the notations of  \cite{Chanfray2007} (except that the screened interactions $V$ are replaced by $G$), the result of the ring summation is
\begin{eqnarray}
E^{ring} &=& E^{ring}_L +E^{ring}_T\nonumber\\
E^{ring}_L&=&\frac{3}{2\rho}\int\frac{id\omega d{\bf q}}{(2\pi)^4} \big[-\ln\big(1-G_{LNN}\Pi^0_N
\nonumber\\
& &-G_{L\Delta\Delta}\Pi^0_\Delta-(G^2_{LN\Delta}-G_{LNN}\,G_{L\Delta\Delta})
\Pi^0_N\Pi^0_\Delta\big)\nonumber\\
& &-G_{LNN}\Pi^0_N-G_{L\Delta\Delta}\Pi^0_\Delta\big]\,\nonumber\\
E^{ring}_T&=&\frac{3}{\rho}\int \frac{id\omega d{\bf q}}{(2\pi)^4}\big[-\ln\big(1-G_{TNN}\Pi^0_N
\nonumber\\
& &-G_{T\Delta\Delta}\Pi^0_\Delta-(G^2_{TN\Delta}-
G_{TNN}\,G_{T\Delta\Delta})
\Pi^0_N\Pi^0_\Delta\big)
\nonumber\\
& &- G_{TNN}\Pi^0_N-G_{T\Delta\Delta}\Pi^0_\Delta\big],
\label{CORREL}
\end{eqnarray} 
where the third lines of the above two  expressions ($E^{ring}_l$, $E^{ring}_T$) for the energy per nucleon correspond to the subtraction of the mean field Fock terms. In short,  $G_L\sim\pi+g'$ and $G_T\sim \rho + g'$ represent  the effective  spin-isospin longitudinal and transverse interaction  in the various $NN$, $N\Delta$ and $\Delta\Delta$  channels. In a  recent paper \cite{Universe}, we used essentially the same approach as in \cite{Chanfray2007} but with
input  parameters appearing in the chiral confining model lagrangian (Eq. \ref{LAG2}).\\
\begin{figure}
  \includegraphics[width=0.3\textwidth,angle=270]{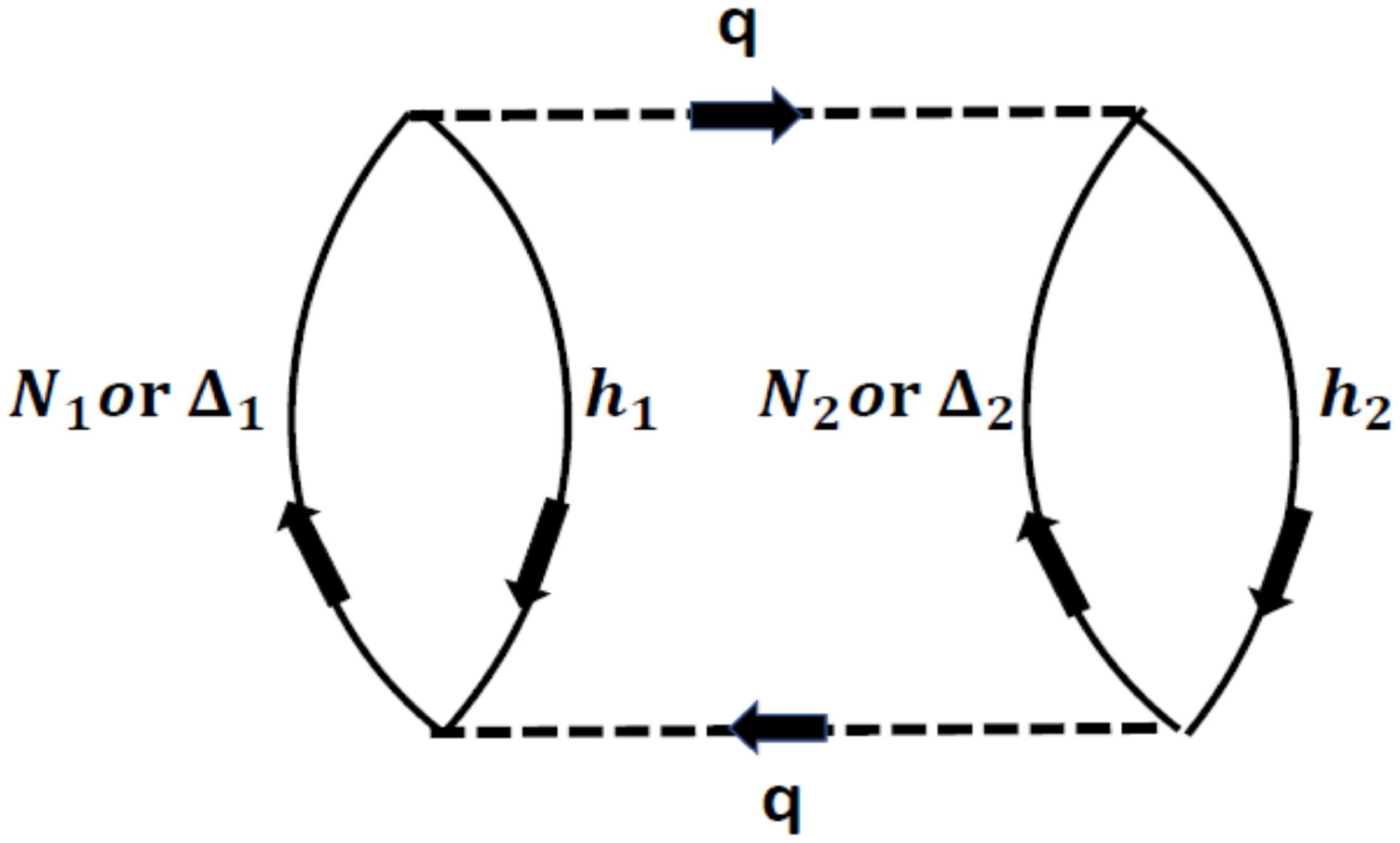}
  \caption{ Diagrams contributing to the  correlation energy or to the depopulation of the Fermi sea.}
  \label{BUBBLE}
\end{figure}

Before going further, let us  summarize the origin of the various input  parameters entering the nuclear matter calculation of  \cite{Universe}  and the new one presented below. For clarity we distinghish those coming from the QCD-connected model from those coming from hadron phenomenology.\\
- For what concerns the parameters entering the bare $NN$ interaction, the scalar and pionic sectors are entirely given or strongly constrained by the QCD- connected model: $g_S=6.52$, $M_\sigma=716.4\,MeV$, $\Lambda_S=1\,GeV$, $g_A=1.26$, $F_\pi=92.4\,MeV$, $m_\pi=140\,MeV$, $\Lambda_\pi=1\,GeV$.  Notice that the cutoff parameters $\Lambda_S,\,\Lambda_\pi $, as well as the cutoff in the vector channel, $\Lambda_V=1\,GeV$, are only approximately  compatible with the QCD- connected model.\\
- The vector-(Lorentz) tensor $NN$ sector is constrained by well established hadron phenomenology: $g_V=7.5$, $m_V=783\,MeV$, $g_\rho=g_V/3$, $m_\rho=770\,MeV$, $\kappa_\rho=6$. We  stress that a large value of $\kappa_\rho$ (the strong rho scenario \cite{HP75} used in the Bonn potential \cite{MHE89,BONNPOT})  is required to decrease the (Wigner) tensor force, to get a not too large D state probability in the deuteron. In the present paper we will use slightly different parameters, namely $g_V=8$ and $\kappa_\rho=6.125$.\\
- The two parameters entering the three-nucleon force, $C=0.32$ and $C_\chi=0.488$,  are given by the QCD-connected model. \\
- Finally the nuclear matter calculation requires a pair correlation function. We used a simple Jastrow ansatz $f(r)=1-j_0(q_c r)$ with $q_c=670\, MeV$ in \cite{Universe} or $q_c=610\, MeV$ in the present paper. The quantity $q_c$, which corresponds to the inverse of the correlation hole size,   is obtained with an adapted G matrix calculation (including the fictitious $\sigma'$, see section 2.3 of Ref. \cite{Universe}) preserving the UV regularization of the loop integrals entering the correlation energy. The effective longitudinal and transverse spin-isospin interactions (up to a factor $g_A^2/4F_\pi^2$ usually included in the definition of the polarization propagator) become $G_L(q)=V_\pi(q) + g'(q)+ 2h'(q)$  and $G_T(q)=V_\rho(q) + g'(q)- h'(q)$ (see the explicit expression of  $g'(q)$ and $h'(q)$  in Eq. (49) of Ref. \cite{Universe}). This represents an efficient way for incorporating the effect of short-range correlations. Moreover, to get a sufficiently large value of the Landau-Migdal parameter ($g'=0.59$ in \cite{Universe}, $g'=0.64$ in the present paper) a rather large value of the cutoff, $\Lambda_\rho=2\,GeV$, for the tensor coupling of the rho meson is needed. This is the only constraint or requirement  from nuclear matter phenomenology \cite{Ichimura}. As a starting point it is a RMF calculation as in \cite{Rahul}, but completed by the inclusion, on top of the Hartree scheme, of the pion and rho loops, namely the Fock energy and the correlation energy, including the contribution of the $\Delta$ resonance. One  main difference with \cite{Chanfray2007} is the utilization of the NJL potential (Eq. \eqref{eq:vchiNJL}) in place of the L$\sigma$M potential (Eq. \eqref{eq:VLSM}) with the value $C_\chi=0.488$ obtained in the underlying QCD-connected NJL model.
Another difference is the replacement of the three purely phenomenological $g'=g'(q=0)$ Landau-Migdal parameters,  by a unique one, $g'_{NN}=g'_{N\Delta}=g'_{\Delta\Delta}=0.59$, generated with the Jastrow ansatz.  As discussed in details in \cite{Universe} we obtained without further fine tuning a decent description of the saturation properties of nuclear matter.
\cite{CEM2022}\cite{CEM2022}\subsection{Correlation energy and Fermi sea depopulation}
This approach is however questionable for various reasons. The first one is the use of a universal $g'$ parameter, i.e, $g'_{NN}=g'_{N\Delta}=g'_{\Delta\Delta}$. There is indeed a consensus that $g'_{NN}$ is larger that $g'_{N\Delta}$ and $g'_{\Delta\Delta}$ and frequently quoted values are $(g'_{NN}, g'_{N\Delta},g'_{\Delta\Delta})=(0.6-0.7, 0.3, 0.5)$ \cite{Ichimura}. As discussed below this deviation from universality  strongly reduces the contribution of diagrams with at least one $\Delta$ in the intermediate state (see fig. \ref{BUBBLE}). In addition this type of iterative $\Delta$ box diagrams, not reducible to the iteration of single meson (pion or rho) exchange in the $NN$ sector,  are often simulated by a fictitious $\sigma'$ meson exchange. In the  $NN$ potential, its contribution, named $2\pi$ in fig. \ref{POTNN}, is obtained using the parameters,
$g_{2\pi}=4.8\equiv g_{\sigma'}$, $m_{2\pi}=550\,\text{MeV}\equiv m_{\sigma'}$, $\Lambda_{2\pi}=1\,\text{GeV}\equiv \Lambda_{\sigma'}$. One second weak point is that the contribution to the binding energy of the $\Delta N$ and $\Delta\Delta$ diagrams depicted in fig. \ref{BUBBLE} is only marginally compatible with the $\sigma'$ contribution. Finally, there is  the old  problem of the double counting in the calculation of the correlation energy on top of the lowest order Brueckner calculation. The two-bubble diagrams of fig. \ref{BUBBLE}  corresponding to the iterated uncorrelated two-pion (and two-rho) exchange, are totally or partially taken into account in the calculation based on a Brueckner G matrix \cite{Friman}. Hence it should be subtracted from the ring diagrams summation.  Moreover,  the explicit calculation  of the diagrams with one or two  $\Delta$'s in the intermediate state, should exactly coincide with the contribution of the $\sigma'$ entering the bare $NN$ interaction, at least   at low density, i.e., in absence of Pauli-blocking effect. To clarify, this question without resorting to a diagrammatic analysis, we reconsider and extend the (non relativistic)  G matrix approach described in section 2 of \cite{Universe}. \\

For this purpose let  us consider a nuclear system described by a Hamiltonian where the bare $NN$ interaction  (i.e., $s,\omega,\pi,\rho$ exchanges) reduces to a two-body potential, written with standard notation as
 \begin{eqnarray}
		H&=& \hat{T} +\hat{V}=\sum_{k_1, k_2}<k_1\,|\,T\,|\, k_2> a^\dagger_{k_1}a_{k_2}\nonumber\\ 
  &+&\frac{1}{4} \sum_{k_1 k_2 k_3 k_4}<k_1\,k_2\,|\,\bar{V}\,|\,k_3\, k_4>
		a^\dagger_{k_1}a^\dagger_{k_2}a_{k_4}a_{k_3},
  \end{eqnarray}
  where  $a$ ($a^\dagger$) is the annihilation (creation) operator relative to the bare nucleons.
According to Goldstone \cite{Goldstone}, the correlated ground state of the nucleus, $|0>$, can be obtained from the state of uncorrelated bare nucleons, $|0>_{uncorr}$, by a unitary transformation as~follows
\begin{eqnarray}
		|0>&=&U(a)\, |0>_{uncorr} \equiv \exp(S(a))  \, |0>_{uncorr}\nonumber\\
         S(a)&=&\frac{1}{4}\left(s_{p_1 p_2 h_1 h_2} a^\dagger_{p_1}a^\dagger_{p_2}a_{h_1}a_{h_2}\,-\,hc\right),
	\end{eqnarray}
with $S(A)$ being an antihermitian operator truncated at 2p-2h excitations (with the notation $s^{*}_{p_1 p_2 h_1 h_2}=s_{h_1 h_2 p_1 p_2}$), which incorporates the effect of short-range correlations. In particular this generates  the Fermi sea depopulation and the occupation number for  holes and particles are \cite{CEM2022}: 
\begin{eqnarray}
		n_h &\equiv& 1-\Delta n^h = 1\,-\,\frac{1}{2}\,\sum_{h_2 p_3 p_4}\left|s_{p_1 p_2 h_1 h}\right|^2\\
		n_p &\equiv& \Delta n^p = \frac{1}{2}\,\sum_{p_2 h_1 h_2}\left|s_{p p_2 h_1 h_2}\right|^2.
	\end{eqnarray}
In addition, notice that, as explained in detail in  \cite{CEM2022}, the $\Delta$ resonance states can be incorporated as extra particle states in all parts of this approach.
To second order in the $s_{p_1 p_2 h_1 h_2}$ parameters the  ground state expectation values of the kinetic energy operator  and of the interaction operator are given  by 
\begin{eqnarray}
\langle\hat{T }\rangle &=&    \sum_h n_h t_h + \sum_p n_pt_p= \langle\hat{T }\rangle_{FG} \nonumber\\
&+&\frac{1}{4}\sum_{h_1 h_2 p_1 p_2}\left|s_{p_1 p_2 h_1 h_2}\right|^2\left(t_{p_1} +t_{p_2}-t_{h_1}-t_{h_2} \right)
\end{eqnarray}
\begin{eqnarray}
\langle\hat{V}\rangle &=& \frac{1}{2}\sum_{h_1 h_2}<h_1 h_2\,|\bar{V}|h_1 h_2> \nonumber\\
&+& \frac{1}{4} \sum_{h_1 h_2 p_1 p_2}<h_1 h_2\,|\bar{V}|p_1 p_2> s_{p_1 p_2 h_1 h_2} \nonumber\\
&+&\frac{1}{4} \sum_{h_1 h_2 p_1 p_2} s_{h_1 h_2 p_1 p_2}<p_1 p_2|\bar{V}|h_1 h_2>\nonumber\\
&+&\frac{1}{4} \sum_{h_1 h_2 p_1 p_2}\left|s_{p_1 p_2 h_1 h_2}\right|^2 \big(v_{p_1}+v_{p_2}-v_{h_1}-v_{h_2}\big) \nonumber\\
&+&\frac{1}{8}\sum_{h_1 h_2 p_1 p_2 p'_1 p'_2}s_{h_1 h_2 p_1 p_2} <p_1 p_2|\bar{V}|p'_1 p'_2>s_{p'_1 p'_2 h_1 h_2} \nonumber\\
&+&\frac{1}{8} \sum_{h_1 h_2 h'_1 h'_2 p_1 p_2}s_{p_1 p_2 h_1 h_2}<h_1 h_2|\bar{V}|h'_1  h'_2>s_{h'_1 h'_2 p_1 p_2}\nonumber\\
\end{eqnarray}
with $t_k=k^2/2 M_N$ for a nucleon state and  $t_{\Delta k}= M_\Delta-M_N + k^2/2M_\Delta$ for a $\Delta$ state. The quantities $v_k=\sum_h <k h|\bar{V}|kh >$ correspond to the HF one-particle potential for the state $k$. We can remark that  we can group together the second line of the kinetic energy and the fourth line of the potential to reconstitute a quantity $$\frac{1}{4} \sum_{h_1 h_2 p_1 p_2}\left|s_{p_1 p_2 h_1 h_2}\right|^2\left(\epsilon_{p_1} +\epsilon_{p_2}-\epsilon_{h_1}-\epsilon_{h_2} \right),$$ where $\epsilon_k=t_k + v_k$
is the single particle energy. In the following we will assume that the one-particle potential is momentum independent, which is true at the Hartree level, so that the $2p-2h$ energy,  $\epsilon_{p_1} +\epsilon_{p_2}-\epsilon_{h_1}-\epsilon_{h_2} $ can be replaced by $t_{p_1} +t_{p_2}-t_{h_1}-t_{h_2}$.
The coefficients $s_{p_1 p_2 h_1 h_2}$can be obtained variationally, i.e., by minimizing the ground state energy, yielding:
\begin{eqnarray}
 && \left(\epsilon_{h_1} +\epsilon_{h_2}-\epsilon_{p_1}-\epsilon_{p_2 }\right)s_{p_1 p_2 h_1 h_2}=<p_1 p_2|\bar{V}|h_1 h_2>\nonumber\\
 && + \frac{1}{2}\sum_{p'_1, p'_2} <p_1 p_2\,|\bar{V}|\,p'_1 p'_2>  s_{p'_1 p'_2 h_1 h_2}\nonumber\\
 &&+\frac{1}{2}\sum_{h'_1, h'_2} s_{p_1 p_2 h'_1 h'_2} <h'_1\,h'_2\,|\,\bar{V}\,|\,h_1\, h_2>.
\end{eqnarray}
This equation has the form of a Feynmann-Galiskii equation. We however neglect the last term involving hole sates which is justified by phase space limitations (see also the discussion around Eqs. 3.8 and 3.9 of the review paper of P. Schuck et al \cite{Schuck89}). In that case the solution of the previous equation can be found as follows:
\begin{equation}
s_{p_1 p_2 h_1 h_2}=   \frac{<p_1\, p_2\,|\,\bar{G}(E=\epsilon_{h_1} +\epsilon_{h_2})\,|\,h_1\, h_2>}{\epsilon_{h_1} +\epsilon_{h_2}-\epsilon_{p_1}-\epsilon_{p_2 }}, \label{eq_spphh}
\end{equation}
where $G(E)$, which can be identified with the Brueckner  G matrix, is a two-body operator satisfying a Bethe-Goldstone equation,
\begin{eqnarray}
 && G(E)=V\,+\, V \,\frac{Q}{E-h_0}\,G(E) \Longleftrightarrow{}\nonumber\\
&&  <k_1\, k_2\,|\,G(E)\,|\,k'_1\, k'_2>=<k_1\, k_2\,|\,V\,|\,k'_1\, k'_2> \nonumber\\
 && + \sum_{p_1, p_2 >k_F} \frac{<k_1k_2|V|p_1 p_2><p_1p_2|G(E)|k'_1 k'_2>}{E-\epsilon_{p_1}-\epsilon_{p_2 }},
\end{eqnarray}
and $Q$ is the Pauli-blocking operator projecting on particle states above the Fermi sea as well as $\Delta$ states. It follows that  he ground state energy per nucleon can be written  as: 
\begin{equation}
 E_0=E^{BHF} + E^{2p-2h}  + E^{depop}.\label{EBHFG}
\end{equation}
$E^{BHF}$ is the mean-field energy associated with the G matrix considered as an  effective interaction, 
\begin{eqnarray}
&&E^{BHF}=\frac{1}{A}\sum_{h<k_F}<h|T|\, h>\nonumber\\
&+&\frac{1}{2A} \sum_{h_1 h_2<k_F}<h_1h_2\,|\bar{G}(E=\epsilon_{h_1} +\epsilon_{h_2})|h_1 h_2>,    
\end{eqnarray}
with
\begin{eqnarray}
&&<h_1h_2\,|\bar{G}(E=\epsilon_{h_1} +\epsilon_{h_2})|h_1 h_2>=<h_1h_2\,|\bar{V}|h_1 h_2>\nonumber\\
&&+\frac{1}{2}\sum_{p_1, p_2 >k_F} \frac{<h_1h_2|\Bar{V}|p_1 p_2><p_1p_2|\Bar{G}(E)|h_1 h_2>}{\epsilon_{h_1} +\epsilon_{h_2}-\epsilon_{p_1}-\epsilon_{p_2 }},\nonumber\\
\end{eqnarray}
where the summation over particle states may also include $\Delta$ states without restriction of momentum.\\
The second term in the expression of the ground state energy \eqref{EBHFG} represents the effect of the G matrix interaction to second order in perturbation theory on top of the mean-field energy: 
\begin{equation}
E^{2p-2h}=  \frac{1}{4A}\sum_{p_1, p_2 >k_F} \frac{\left|<p_1p_2|\Bar{G}(E)|h_1 h_2>\right|^2}{\epsilon_{h_1} +\epsilon_{h_2}-\epsilon_{p_1}-\epsilon_{p_2 }}.  
\end{equation}
To make an explicit connection with the ring diagrams energy,  let us come back to the correlation energy to the two-loop (two-bubble or two-hole-line) order:
\begin{eqnarray}
E^{ring-2h} &=&E^{ring-2h}_L\,+\, E^{ring-2h}_T\nonumber\\
E^{ring-2h}_L
&=&\frac{3}{2\rho}\int \frac{id\omega d{\bf q}}{(2\pi)^4}\,\big[G^2_{LNN}\Pi^{02}_N\,+\,G^2_{L\Delta\Delta}\Pi^{02}_\Delta\nonumber\\
&& \qquad\qquad + 2\,G^2_{LN\Delta}\Pi^0_N\Pi^0_\Delta\big]\nonumber\\
E^{ring-2h}_T
&=&\frac{3}{\rho} \,\int\frac{id\omega d{\bf q}}{(2\pi)^4}\,\big[G^2_{TNN}\Pi^{02}_N\,+\,G^2_{T\Delta\Delta}\Pi^{02}_\Delta\big]\nonumber\\
&&\qquad\qquad + 2\,G^2_{TN\Delta}\Pi^0_N\Pi^0_\Delta\big]
.\label{2p-2h}
\end{eqnarray}
If we remove the energy dependence of the spin-isospin interaction, i.e., in a static approximation, one can show that this two-loop correlation energy takes the simple form:
\begin{eqnarray}
E^{ring-2h} &\simeq & E^{corr-st}\qquad\hbox{(static limit)}\nonumber\\
E^{corr-st}&=&\frac{1}{2A}\,\sum_{h_1,h_2, p_1, p_2}\frac{\left|<p_1\, p_2\,|\,G_{\sigma\tau}\,|\,h_1\, h_2>\right|^2}{\epsilon_{h_1}+\epsilon_{h_2}-\epsilon_{p_1}-\epsilon_{p_2}},\label{ESTATIC}
\end{eqnarray}
where $G_{\sigma\tau}$ is the static spin--isospin interaction. We see that $E^{corr-st}$
 coincides with the 2p-2h energy, $E^{2p-2h}$, but omitting the Pauli exchange term which does not appears in the ring summation.\\ Finally the last term appearing in the expression of the BHF ground state energy \eqref{EBHFG} is: 
\begin{eqnarray}
E^{depop}&=&\frac{1}{A}\sum_h (1-n_h)\epsilon_h  + \sum_p n_p\epsilon_p\nonumber\\
&\equiv& \frac{1}{4A}\sum_{p_1, p_2 >k_F} \frac{\left|<p_1p_2|\Bar{G}(E)|h_1 h_2>\right|^2}{\epsilon_{p_1} +\epsilon_{p_2}-\epsilon_{h_1}-\epsilon_{h_2 }}\nonumber\\
&=& -E^{2p-2h} \simeq T^{depop}= \frac{1}{A}\langle T\rangle_{depop}.
\end{eqnarray}
Hence we see that the leading order (two-bubble or two-hole-line) correlation energy, $ E^{ring-2h}\simeq E^{corr-st}$, identified with the 2p-2h energy, $E^{2p-2h}$,  is just compensated by the depopulation energy, $E^{depop}$, itself very close to the excess of kinetic energy per nucleon due to short range correlations. Said differently one has: 
$$E^{ring-2h}\simeq E^{corr-st}\simeq - \frac{1}{A}\langle T\rangle_{depop}.$$ 
This excess of kinetic energy per nucleon not larger than $20\,MeV$ brings a very strong constraint on this correlation energy and especially on the magnitude of the loop energy involving $\Delta$'s in the intermediate state. As emphasized in \cite{Dewulf}, this question of the excess of kinetic energy, associated with the tail of the genuine hole spectral function,  is related to the impact of short-range correlations on the binding energy of nuclear matter according to the well-known  non relativistic energy sum rule,  
\begin{eqnarray}
B&=&		\frac{2}{\rho}\int dE\int\frac{d{\bf k}}{(2\pi)^3}\left(\frac{k^2}{2M}\,+\, E \,S^h(E,{\bf k})\right)\nonumber\\
&\equiv&\frac{1}{2}\left(\frac{1}{A}\langle T\rangle\,+\,\epsilon\right),
	\end{eqnarray}	
where the second form is known as the Koltun sum rule. $S^h(E,{\bf k})$ is the hole spectral function whose explicit form in our approach is given by Eq. (24) of \cite{CEM2022}. $\epsilon$ is the separation energy, a negative quantity which is the opposite of the mean energy needed to remove a nucleon from the nucleus. Let us consider the case  $\rho=0.8 \,\rho_0$, i.e., $k_F=245\,MeV$, appropriate for the carbon nucleus. The results are extremely sensitive to the parameters used for the $N\Delta \rho$ couplings. With the choice discussed below, yielding $(g'_{NN}, g'_{N\Delta}, g'_{\Delta\Delta})$ $=(0.64, 0.30, 0.48)$,  we find $E^{ring-2h}=-20\,MeV$ and $E^{corr-st}=-16\,MeV$, the difference between the non static and the static calculations coming from the longitudinal channel. This is expected because the static pion exchange is certainly a poor approximation due to the smallness of the pion mass. However the effect of Pauli exchange is also expected to decrease this correlation energy and we can take $\langle T/A\rangle_{depop}=- E^{corr-st}=16\,MeV$, with a contribution of about $5\,MeV$ coming from $\Delta$'s present in the ground state. It follows that $\langle T/A\rangle\simeq 20 + 16 = 36\,MeV$ and $-\epsilon =-2B +\langle T/A\rangle\simeq 16 +36 \simeq 50\,MeV$. This is within the estimate obtained with variational calculations \cite{Ciofi}, which is needed to explain the EMC effect \cite{EMC84} (see also the discussions given in \cite{CEM2022}). It is also interesting to have an estimate of the mean depopulation  of the Fermi sea   also calculated with the neglect of Pauli exchange,
\begin{equation}
		\Delta n=\frac{1}{2A}\,\sum_{h_1,h_2, p_1, p_2}\left|\frac{<p_1\, p_2\,|\,G\,|\,h_1\, h_2>}{\epsilon_{h_1}+\epsilon_{h_2}-\epsilon_{p_1}-\epsilon_{p_2}}\right|^2  \simeq 0.175,\label{NSTATIC}
\end{equation}
which is also in agreement with \cite{Ciofi}.
 \subsection{the "$\sigma'$ meson" and the $\Delta$ box diagrams}
Let us  come back to the two-nucleon problem. As already explained the G matrix has been generated from a Bonn-like potential, $V$ , containing genuine  $s$, $\omega$, $\rho$ and $\pi$ exchanges appearing  in the model lagrangian  (Eq. \eqref{LAG2}) and complemented by a $\sigma'$ exchange simulating the iterative isobar box diagrams as in \cite{Holinde}. We write this effective OBE quasi potential  as $V^{(QP)}=V+V_{\sigma'}$. Hence the Bethe-Goldstone equation projected onto the nucleon sector  reads with schematic notations 
\begin{eqnarray}
G_{NN}&=& V_{NN}^{(QP)}\, +\, V_{NN}^{(QP)}\, \left(\frac{Q}{e}\right)_{NN}\, G_{NN}\nonumber\\   
 &\equiv& \,V_{NN}^{(QP)} \,\Omega, \label{GQP}
\end{eqnarray}
with $\Omega=1+\left(\frac{Q}{e}\right)_{NN}\, G_{NN}$.
In our recent paper \cite{Universe} we have shown that  the M\"{o}ller operator $\Omega$ can be represented by a unique Jastrow function, $f_c(r)=1- j_0 (q_c r)$, which plays the role of a unique state independent correlation function as in the LOCV variational approach (see a detailed discussion of this point in Ref. \cite{Baldo2012}). Since $f_c(r=0)=0$, an immediate consequence is that the effective interaction automatically vanishes at high momentum providing a natural regularization of the loop integrals entering for instance the two-pion (or two-rho) exchange diagrams. As discussed in \cite{Chanfray1985}, this property can be seen as a consequence of the Beg-Agassi-Gal theorem which states that the virtual mesons propagate freely inside the correlation hole. Of course such a prescription yields a G matrix which cannot be an exact solution the Bethe-Goldstone equation but only an approximate one. In \cite{Universe} we proposed a method to find the value of $q_c$ preserving the property $f_c(r=0)=0$ after one G matrix iteration. We now require that the G matrix obtained with the fictitious $\sigma'$ is equivalent at least at low density with the the one obtained after the explicit incorporation of  $\Delta$'s in the intermediate state. For this purpose we write the Bethe-Goldstone equation \eqref{GQP}  in  two equivalent forms, one involving the quasi potential $V^{(QP)}=V+V_{\sigma'}$, without $\Delta$ states, and  the other one involving the genuine $OBE$ potential $V$\, with explicit incorporation of the $\Delta$ states (fig. \ref{SIGEXCH}b):
\begin{eqnarray}
G_{NN}&=& V_{NN}+  V_{NN}\, \left(\frac{Q}{e}\right)_{NN}\, G_{NN}\nonumber\\
&& + \,V_{\sigma'} +V_{\sigma'}\left(\frac{Q}{e}\right)_{NN}\, G_{NN} 
\nonumber\\
&=&V_{NN} \Omega \,+\,V_{\sigma'}\Omega\equiv \,G_{NN}^{(OBE)}\,+\,G_{\sigma'},
\end{eqnarray}
\begin{eqnarray}
 G_{NN}&=& V_{NN}+  V_{NN}\, \left(\frac{Q}{e}\right)_{NN}\, G_{NN}\nonumber\\
 &&\,+ V_{N\Delta}\, \left(\frac{Q}{e}\right)_{N\Delta}\, G_{N\Delta}\nonumber\\
&&\,+  V_{\Delta\Delta}\, \left(\frac{Q}{e}\right)_{\Delta\Delta}\, G_{\Delta\Delta}\nonumber\\
&=& V_{NN}+  V_{NN}\, \left(\frac{Q}{e}\right)_{NN}\, G_{NN}\, + \,G^{(\Delta \,box)}_{NN}\nonumber\\
&\equiv& \,G_{NN}^{(OBE)}\, + \,G^{(\Delta \,box)}_{NN}.
\label{GDELTA}\end{eqnarray}
Hence we arrive at the conclusion that the effective interaction $G_{\sigma'}=V_{\sigma'}\Omega$ (i.e., $G_{\sigma'}(r)=V_{\sigma'}(r)(1-j_0(q_cr))$ should be equivalent to the one generated by the iterative $\Delta$ diagrams corresponding to the second and third  lines of Eq. (\eqref{GDELTA}) and depicted in fig. \ref{SIGEXCH}b. This equivalence should be exact in the low density limit in order to generate an identical 
scattering T matrix. We  now adjust the $N\Delta$ and the $\Delta\Delta$ interactions in such a way that the binding energy calculated with the auxiliary "$\sigma'$ meson"  is the same as the one issued from the iterative $\Delta$ box diagrams. In other words we require in the low density limit:
\begin{eqnarray}
E_H^{\sigma'}&=&    \frac{1}{2A}\,\sum_{h_1,h_2}<h_1\, h_2\,|\,G_{\sigma'}\,|\,h_1\, h_2>\nonumber\\
&=&\frac{1}{2A}\,\sum_{h_1,h_2}<h_1\, h_2\,|\,G^{(\Delta \,box)}_{NN}\,|\,h_1\, h_2>\nonumber\\
&=&
\frac{1}{2A}\,\sum_{h_1,h_2; p_1, p_2}<\,h_1\, h_2|\,V_{\sigma\tau}\,|(p_1\, p_2)_\Delta>\nonumber\\
& &\frac{<(p_1\, p_2)_\Delta|\,G_{\sigma\tau}\,|\,h_1\, h_2>}{\epsilon_{h_1}+\epsilon_{h_2}-\epsilon_{p_1}-\epsilon_{p_2}}
=E^{\Delta\, box}.
\end{eqnarray}
We now adapt these  non relativistic results to our relativistic lagrangian approach. For this purpose we introduce an auxiliary lagrangian, $\mathcal{L}_{\sigma'}=-g_{\sigma'}\Bar{N}\sigma' N+\partial^\mu\sigma'\partial_\mu\sigma'/2 -m^2_{\sigma'}\,{\sigma'}/2$, with $g_{\sigma'}=4.8$ and $m_{\sigma'}$=550\,MeV, completed by a monopole $\sigma'NN$ vertex form factor with cutoff $\Lambda_{\sigma'}=1\,GeV$. The corresponding  Hartree energy in presence of short range correlation reads:
\begin{eqnarray}
E_H^{\sigma'}&=&- \frac{g^2_{\sigma'}}{2\,m^2_{\sigma'}}\frac{\rho^2_S}{\rho} \left(1-\Lambda_{\sigma'}^2(q^2_c)\frac{  m^2_{\sigma'}}{m^2_{\sigma'}+ q^2_c}\right)\nonumber\\
&&+\frac{g^2_{\sigma'}}{m^2_{\sigma'}}\frac{\rho^2_S-\rho^2}{\rho}.\label{EHPRIME}
\end{eqnarray}
The iterative $\Delta$ box diagram is calculated using non static $\pi$   and $\rho$ exchanges: 
\begin{eqnarray}
E_H^{\Delta\, box}&=&  E_H^{L\Delta\, box}+E_H^{T\Delta\, box}\nonumber\\
E_H^{L\Delta\, box}&=&\frac{3}{2\rho}\int \frac{id\omega d{\bf q}}{(2\pi)^4}\,\big[2\,V_{\pi N\Delta}\,G_{LN\Delta}\,\Pi^0_N\Pi^0_\Delta\nonumber\\
&& \qquad\qquad + V_{\pi\Delta\Delta}\,G_{L\Delta\Delta}\Pi^{02}_\Delta \big]\nonumber\\
E_H^{T\Delta\, box}
&=&\frac{3}{\rho} \,\int\frac{id\omega d{\bf q}}{(2\pi)^4}\,\big[2\,V_{\rho N\Delta}\,G_{TN\Delta}\,\Pi^0_N\Pi^0_\Delta\nonumber\\
&&\qquad\qquad + V_{\rho\Delta\Delta}\,G_{T\Delta\Delta}\Pi^{02}_\Delta \big]
.\label{EHDEL}
\end{eqnarray}    
The  effective spin-isospin interactions in the $N\Delta$ and $\Delta\Delta$ channels are obtained from the bare interactions with the same  M\"{o}ller operator, i.e, $G_{N\Delta}=V_{N\Delta}\Omega$ and $G_{\Delta\Delta}=V_{\Delta\Delta}\Omega$. In order to  satisfy the constraint $E_H^{\sigma'}=E_H^{\Delta\, box}$ together with the  phenomenological requirement of having different  $g'$ parameters, we take different $\rho$ meson form factors  in the $N\Delta$ and $\Delta\Delta$ channels. If we choose $\Lambda_{\rho N\Delta}=600\,MeV$ and $\Lambda_{\rho \Delta\Delta}=1041\,MeV$, we obtain $(g'_{NN}, g'_{N\Delta}, g'_{\Delta\Delta})$ $=$
$(0.64, 0.30, 0.48)$ which is very close to what is suggested in \cite{Ichimura}. 
\begin{figure}
 \centering
  \includegraphics[width=0.5\textwidth,angle=0]{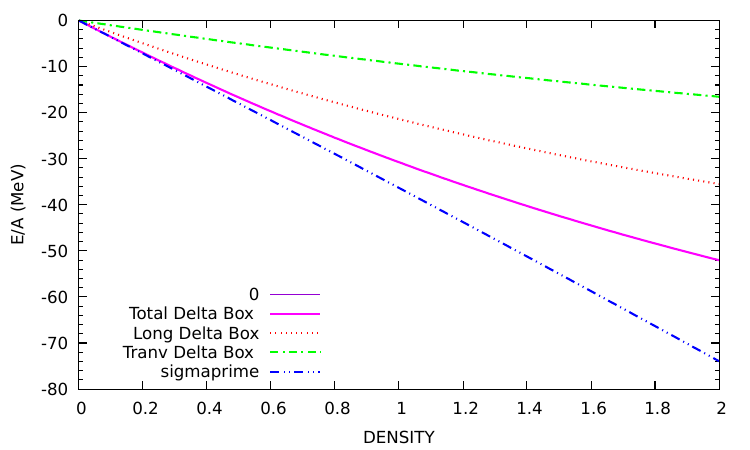}
  \caption{ Contribution to the energy per nucleon versus $\rho/\rho_0$, with $\rho_0=0.16\,fm^{-3}$, of the $\Delta$ box diagrams compared with the exchange  of the fictitious sharp  $\sigma'$ meson. Full line:  total iterative $\Delta$ box. Dotted line: longitudinal $\Delta$ box. Dotted dashed line: transverse $\Delta$ box. Double dotted-dashed line: Hartree $\sigma'$ exchange. }
  \label{Sigmaprime}
\end{figure}
In fig. \ref{Sigmaprime}, we show the comparison between the $\Delta$ box energy (depicted in fig. \ref{SIGEXCH}b) and the Hartree $\sigma'$ energy (depicted in fig. \ref{SIGEXCH}a). We see that with increasing density these $\Delta$ diagrams are less and less well represented by this fictitious $\sigma'$ which lacks the Pauli blocking effect present in the p-h bubble of fig. \ref{SIGEXCH}b1. This is the origin of the genuine repulsive three-body force depicted in fig. \ref{THREEBODY}c corresponding in a different language 
to the earlier celebrated forces proposed  in Refs. \cite{TB1,TB2,TB3,TB4} and also contained at NNLO in chiral EFT \cite{CEFT}. We also see in fig. \ref{Sigmaprime} that the longitudinal spin-isospin contribution (iteration of the pion exchange) is twice larger than the transverse one (iteration of the rho exchange) hence justifying the usual terminology of "two-pion exchange" simulated by the scalar-isosclar $\sigma'$ exchange.
\subsection{The nuclear matter equation of state}
In our previous paper \cite{Universe}, the G matrix was actually generated starting from a bare potential keeping only in the spin-isospin component its central part, 
$$V_{C}({\bf{q}})=\frac{1}{3}\big(V_\pi(q) + 2V_\rho(q)\big){\bf{\sigma}}_1 \cdot{\bf{\sigma}}_2, $$  ignoring the "weak" tensor force:
$$V_W({\bf{q}})=\frac{1}{3}\big(V_\pi(q) -V_\rho(q)\big)\left(3{\bf{\sigma}}_1\cdot\hat{\bf{q}}\,{\bf{\sigma}}_2 \cdot\hat{\bf{q}}\, -\,{\bf{\sigma}}_1\cdot{\bf{\sigma}}_2\right).$$
Hence if we write the quasi potential, including the $\sigma'$, as $V^{(QP)}=V_S+V_W$, the Bethe-Goldstone equation which was actually solved in \cite{Universe} reads
\begin{equation}
G_S= V_S\,+ \, V_S\, \left(\frac{Q}{e}\right)_{NN}\, G_S,    
\end{equation}
with solution $G_S=V_S\,\Omega_S$. $\Omega_S$ corresponds to the aforementioned M\"{o}ller operator represented by a unique Jastrow function, $f_c(r)=1-j_0(q_c(r)$ (i.e., $G_S(r)=V_S(r)$ $f_c(r)$), whereas the "exact"  Bethe-Goldstone equation should be:
\begin{equation}
G= V_S+ V_W +(V_S+V_W)\, \left(\frac{Q}{e}\right)\, G.   
\end{equation}
where we  now omit the suffix $NN$ in the Pauli operator. 
This equation can be rewritten in the successive forms  as:
\begin{eqnarray}
G&=&V_S+V_S \left(\frac{Q}{e}\right)\, G_S \,+\,  V_W+V_W \left(\frac{Q}{e}\right)\, G_S \nonumber\\
&+&(V_S+V_W) \left(\frac{Q}{e}\right) (G-G_S)\nonumber\\
&=& G_S + V_W\Omega_S\, +\,(V_S+V_W) \left(\frac{Q}{e}\right) (G-G_S)\\
&\simeq& G_S+G_W +(V_S+V_W) \left(\frac{Q}{e}\right) G_W\\
&=& V^{(QP)} \Omega_S+ (V_S+V_W) \left(\frac{Q}{e}\right) G_W.\label{TENSOR}
\end{eqnarray}
To obtain the last form we have approximated the difference $G-G_S$ by $G_W=V_W\Omega_S$, ignoring terms such as: $$V^{(QP)}\left(\frac{Q}{e}\right) V^{(QP)}\left(\frac{Q}{e}\right)G_W+...\, .$$ $G_W$ is the effective interaction in the tensor channel in presence of short-range correlations obtained from the bare tensor interaction according to 
$$
\frac{1}{3}\big(V_\pi(q) -V_\rho(q)\big)\quad\to\quad \frac{1}{3}\big(V_\pi(q) -V_\rho(q) +3 h'(q)\big) ,  
$$
$h'(q)$, as well as $g'(q)$, being given by Eq. (49) of \cite{Universe}.
The first term of Eq. \eqref{TENSOR} represents the full G matrix obtained 
from the scalar-isoscalar $s=\sigma_W$ exchange, the  $\omega$ exchange,  the Lorentz-vector piece of $\rho$ exchange and the full spin-isospin $\pi+\rho$ exchange including the  tensor force, using the Jastrow ansatz. In the second term of Eq. \eqref{TENSOR},  only the $V_W (Q/e)G_W$ term survives after the spin-isospin trace. It follows that  the G matrix in the NN sector, incorporating the iteration of the tensor potential, reads:
\begin{equation}
G= G_S +G_W +V_W \left(\frac{Q}{e}\right) G_W.   
\end{equation}
In the calculation of the binding energy the BHF contribution of $G_W$,
\begin{equation}
E^{BHF-tensor}=  \frac{1}{2A}\sum_{h_1,h_2}<h_1 h_2|\bar G_W|h_1 h_2>\equiv 0,
\end{equation}
identically vanishes for either the direct term or the exchange term. Hence the spin-isospin tensor potential contributes to the binding energy only through the second term of Eq. \eqref{TENSOR}. Relaxing the static limit, this tensor loop contribution writes: 
\begin{equation}
E^{tensor}=\frac{1}{\rho}\int \frac{id\omega d{\bf q}}{(2\pi)^4}\,(V_\pi-V_\rho)(G_{LNN}-G_{TNN})\Pi^{02}_N.\label{TENSORLOOP}
\end{equation}
For completenes we recall that this type of  calculation, such as Eqs. \eqref{CORREL},\eqref{2p-2h},\eqref{TENSORLOOP},  fully incorporates the recoil correction and the energy dependence of the spin--isospin interaction, which means that the $q^2$ appearing in the pion and rho  static propagators  is systematically replaced by $q^2 -\omega^2$, and the integral $\int id\omega d{\bf q}$ becomes $-\int dz d{\bf q}$ after a Wick rotation (see \cite{Chanfray2007,Massot2009}), hence transforming $q^2 -\omega^2$ into $q^2 +z^2$.
\\

We are now in position to calculate the nuclear matter equation of state.  The total energy per nucleon can be decomposed according to:
\begin{equation}
E_0=E_0^{RHF} + E^{\pi\rho-Fock}+ E^{2\pi-loop}.
\end{equation}
** $E_0^{RHF}$ corresponds to the mean-field RBHF result with, $s$, $\omega$ and  (Lorentz vector piece of) $\rho$ exchanges. In this calculation, as in \cite{Universe}, the nucleon scalar self-energy also includes the effect of the $\sigma'$, improving the agreement  with the Hugenholtz-Van-Hove theorem and yielding an effective Dirac mass, $M^*_N(\rho_0)=764 \,MeV$.  It contains the three-body forces originating  from the nucleon response to the scalar field (fig. \ref{THREEBODY}a) and from the chiral tadpole diagram (fig. \ref{THREEBODY}b) whose net effect is to generate a repulsive three-body contribution to the energy per nucleon given by Eq. \eqref{eq:THREEBODN}.  It also includes the Fock terms but with an averaged value of the exchanged momentum (see Eqs. (117,118) of \cite{Universe}). We  also have to mention that this Fock term is included perturbatively in the sense that we remain in the Hartree basis, since we neglect the Fock contribution to the scalar self-energy of the nucleon. Although this problem is of minor importance for the binding energy around the nuclear matter density, the use of a fully self-consistent Hartree--Fock basis in place of the Hartree basis for the nucleon Dirac wave function may play a very important role at a high density in limiting the maximum mass of a hyperonic neutron star, as pointed out in Ref.~\cite{Massot2012}. The passage to this Hartree--Fock basis can be implemented without formal difficulty  but at the expense of increasing  numerical complexity \cite{Massot2008,Cham}. We also introduce the effect of short-range correlations by subtracting from the Hartree and Fock contributions the same quantities but  adding $q^2_c$ to the exchange momentum squared in the meson propagators and form factors.\\
** $E^{\pi\rho-Fock}$ is the Fock term associated with the $\pi+\rho$ spin-isospin intetaction in presence of short-range correlations. Notice that only the central part of the spin-isospin interaction gives an non vanishing contribution; in particular  it involves only g'(q) but not $h'(q)$.\\
** The two-pion (or two-rho) loop contribution is the sum of three pieces:
\begin{equation}
E^{2\pi-loop}= E^{corr-3h} \,  +  \,E^{\Delta\, box}\,+\,E^{tensor}.
\end{equation}
* $E^{corr-3h}= E^{ring}-E^{ring-2h}$ is the genuine correlation energy  starting at the three-hole-line order which is obtained by subtracting  the two-bubble (two-hole-line) 2p-2h energy, $E^{ring-2h}$  \eqref{2p-2h}, from the full ring summation, \eqref{CORREL}. As demonstrated above, the reason is the depopulation of the Fermi sea due to short-range correlations, inducing an excess of kinetic energy which just compensates this 2p-2h energy,  $E^{ring-2h}$.\\
* $E^{\Delta\, box}$ is the contribution of the iterative $\Delta$ box diagrams depicted in fig.\ref{SIGEXCH}b. Since one has by construction at the hartree level $E_H^{\Delta\, box}\simeq E_H^{\sigma'}$, we incorporate the Pauli exchanged  term according to 
\begin{equation}
E^{\Delta\, box} =E_H^{\Delta\, box}\left(1-\frac{E_F^{\sigma'}}{E_H^{\sigma'}}\right),   
\end{equation}
with $E_H^{\Delta\, box}$ given by \eqref{EHDEL} and $E_H^{\sigma'}$ given by \eqref{EHPRIME}. The Fock $\sigma'$ energy is computed as
\begin{eqnarray}
E_F^{\sigma'}&=& \frac{g^2_{\sigma'}}{16}\,\frac{\rho^2\,+\,\rho_S^2}{\rho}   
\left(\frac{\Lambda_{\sigma'}^2(\bar{q}^2)}{m^2_{\sigma'}+\bar{q}^2}-\frac{\Lambda_{\sigma'}^2(\bar{q}^2+q^2_c}{m^2_{\sigma'}+\bar{q}^2 +q_c^2}\right),
\end{eqnarray}
with $\bar{q}^2=6 k_F^2/5$.\\
* $E^{tensor}$ given by Eq. \eqref{TENSOR} is the contribution of the iterated $\pi-\rho$ tensor force in presence of short-range correlations.

\begin{figure}
 \centering
  \includegraphics[width=0.5\textwidth,angle=0]{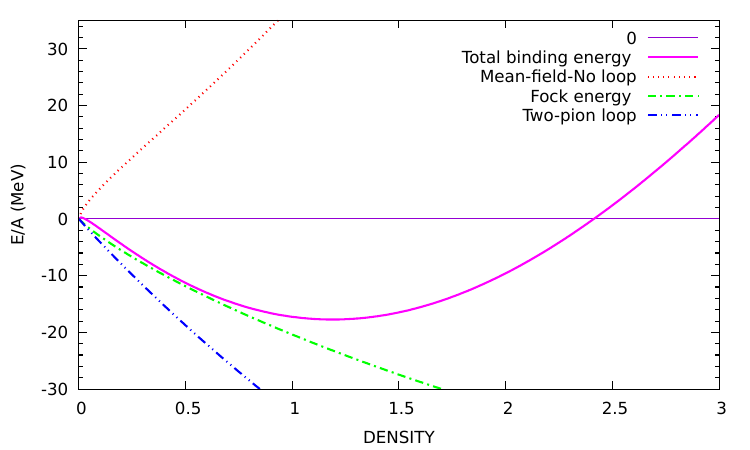}
  \caption{ Binding energy of nuclear matter versus $\rho/\rho_0$. Full line: total energy. Dotted line: mean field RHF result. Dot-dashed line: spin-isospin Fock term. Double dotted-dashed line: $2\pi$-loop.}
  \label{EOSFINAL}
\end{figure}

\begin{figure}
 \centering
  \includegraphics[width=0.5\textwidth,angle=0]{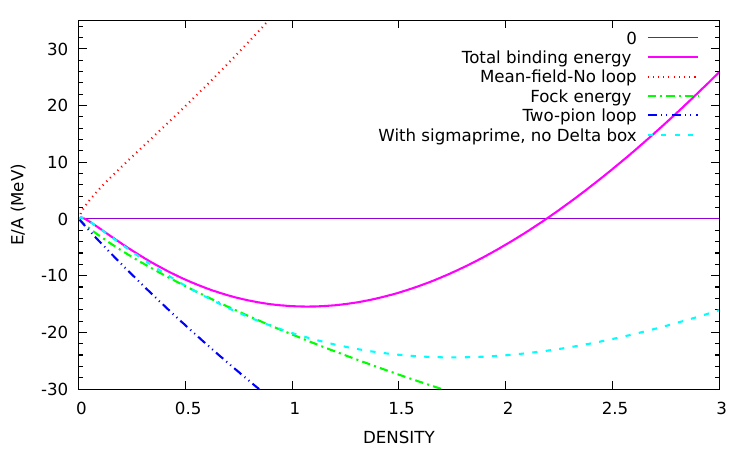}
  \caption{ the same as fig. \ref{EOSFINAL} but with $C=0.42$ in place of $C=0.32$. In addition we also show the result of the calculation when the $\Delta$ box diagrams are replaced by the $\sigma'$ exchange (dashed line).}
  \label{EOSFINALb}
\end{figure}

Even if we are aware that many of the various approximations or prescriptions of this multi-step approach should or would deserve a more detailed study, we do not refrain from showing  the results of the calculation without a further fine-tuning of the parameters.  The saturation curve is displayed in Fig. \ref{EOSFINAL}. The binding energy at the saturation point, $E_{sat}/A=-17.73\,MeV$, and the saturation density, $\rho=1.19\,\rho_0=0.19\,{fm}^{-3}$, are two large, as well as the incompressibility modulus, $K_{sat}=338\, MeV$. Even though the saturation point is not perfect, it is rather satisfactory to arrive at decent results for the properties of nuclear matter in an approach based on a microscopic model that mainly  depends on QCD inputs, namely, the string tension and the gluon correlation length (or the gluon condensate) and parameters ($g_V/m_V, \,\kappa_\rho$) known from well-established hadronic phenomenology.\\
It is not the purpose of this article to adjust the parameters derived from the QCD-connected model, but we simply notice that if we increase the $C$ parameter from $C=0.32$ to $C=0.42$, we can improve the results for the saturation point: $E_{sat}/A=-15.45\,MeV$, $\rho=1.07\,\rho_0=0.17\,\text{fm}^{-3}$, $K_{sat}=290\, MeV$ (Fig. \ref{EOSFINALb}), whereas the lattice parameter $a_4= -0.28\,GeV^{-3}$ is compatible with lattice results. A robust conclusion is that the pion and  rho loops  are necessary to obtain sufficient binding, a conclusion already reached in Ref. \cite{Chanfray2007}, even if this contribution is much  smaller than what is obtained
from iterated pion exchange (planar diagrams) in in-medium chiral perturbation theory \cite{KFW02} where  the short-range physics is accounted for by a unique cutoff regularizing the pion loops.
\subsection{ The two "sigma mesons" and their associated three-body forces} 
\begin{figure}
  \includegraphics[width=0.5\textwidth,angle=0]{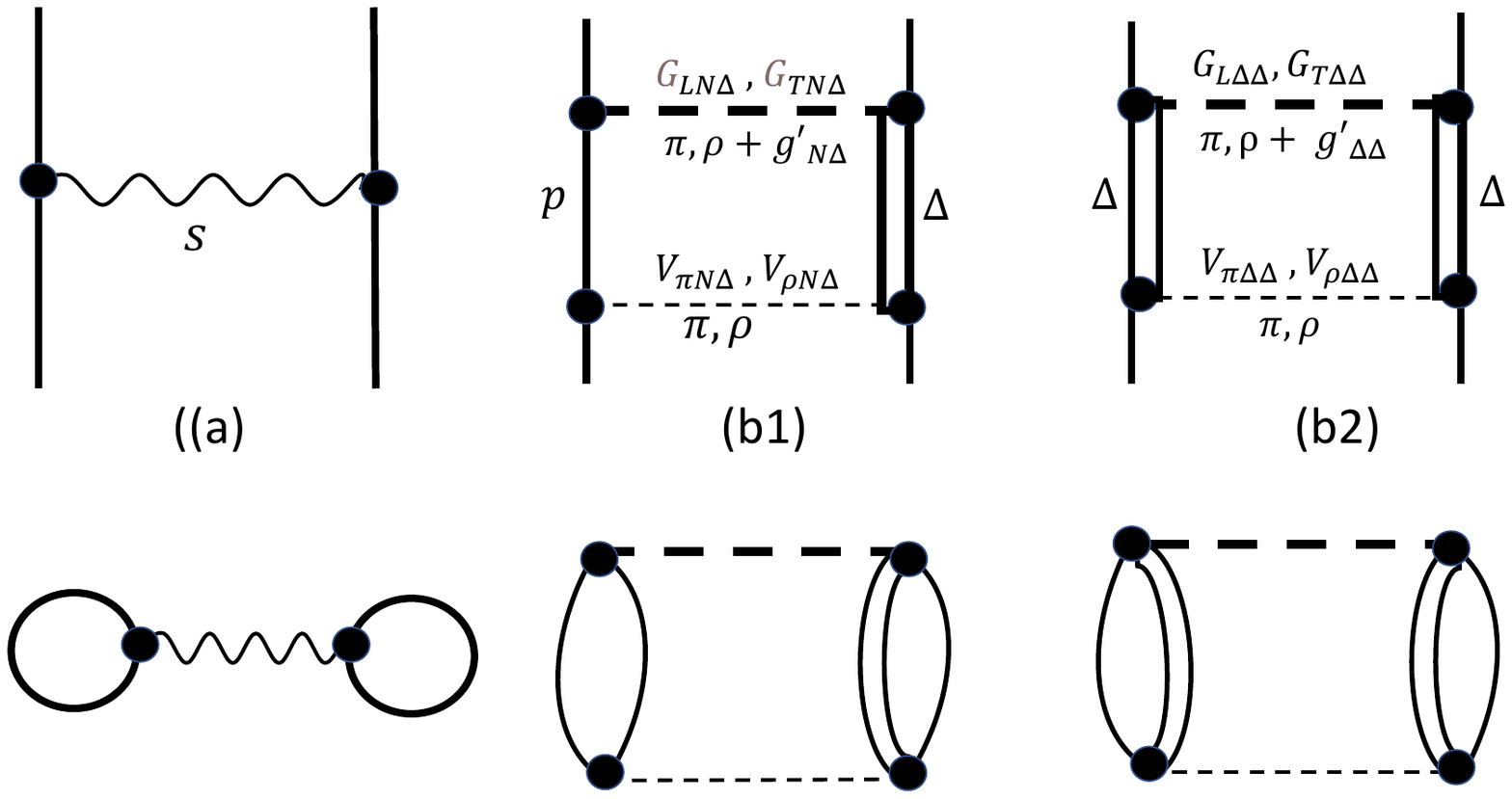}
  \caption{Upper panel: diagrams contributing to the scalar-isoscalar interaction in nuclear matter. (a) hiral scalar field, $\sigma_W=s$, exchange; (b) $\Delta$ box- $2 \pi$ exchange, often simulated  by  a fictitious $\sigma'$. Lower panel: the equivalent many-body diagrams.}
  \label{SIGEXCH}
\end{figure}

\begin{figure}
  \includegraphics[width=0.48\textwidth,angle=0]{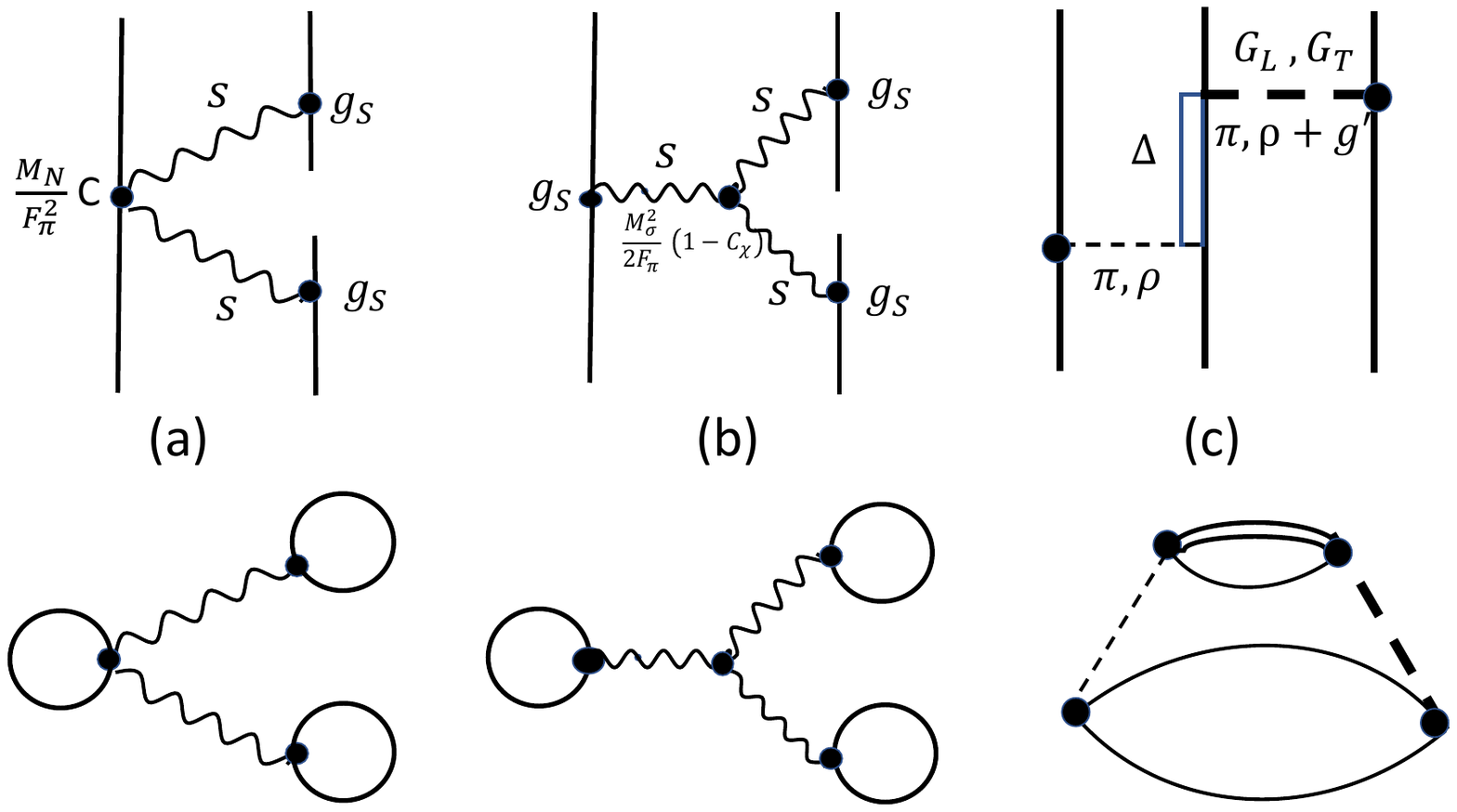}
  \caption{Upper panel:  Three-body forces (3-hole-line) associated with scalar-isoscalar exchange in nuclear matter. (a) response of the nucleon to the scalar field $s$ generating a repulsive three-body force;  (b) tadpole diagram from the $s^3$ term in the chiral effective potential generating an attractive three-body force; (c) in-medium modification of the $\pi\rho$  Fock term from Pauli-blocking in diagram (b1) of fig. \ref{SIGEXCH} ; this diagram generating a repulsive three-body force can also be seen as a in-medium modification of the $\sigma'$ meson due to Pauli-blocking. Lower panel: the equivalent many-body diagrams. }
  \label{THREEBODY}
\end{figure}

\begin{figure}
  \includegraphics[width=0.48\textwidth,angle=0]{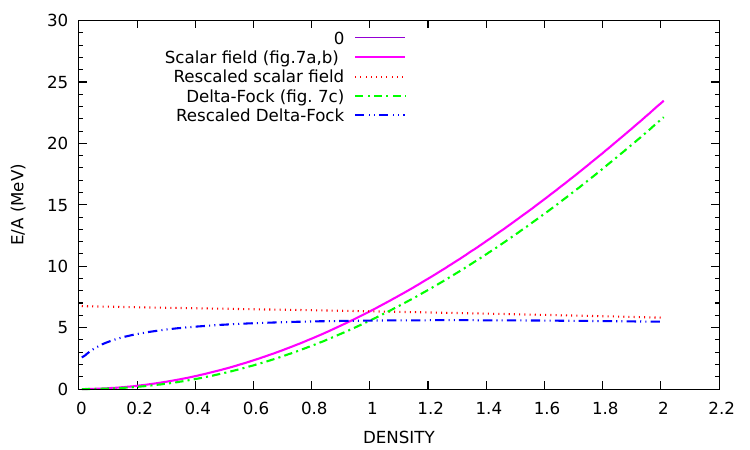}
  \caption{ Contribution to the binding energy per nucleon coming from the three-body diagrams of the lower panel of fig. \ref{THREEBODY}, versus $\rho/\rho_0$. Full line:  three body force (nucleon response + tadpole) associated with the chiral scalar field $s$ (fig. \ref{THREEBODY}a,b); dotted line: the same but rescaled by $\rho^2_0/\rho^2$; dot-dashed line: correction to the spin-isospin Fock-term by $\Delta-h$ bubble dressing (fig. \ref{THREEBODY}c); double dotted-dashed line: the same but rescaled by $\rho^2_0/\rho^2$ .}
  \label{THREEBODYCC}
\end{figure}

I now come back to the questions raised in our earlier works of the nineties about the status of the sigma meson in nuclear physics. The conclusion is that the scalar-isoscalar attraction has two quite different origins associated in some sense with two "sigma mesons". The first one is associated with the radial fluctuation of the chiral condensate in the chirally broken vacuum (fig. \ref{SIGEXCH}a). In the field theoretical  lagrangian approach, this mode corresponds to the scalar-isoscalar field, $s=\sigma_W$, and can be identified with the "nuclear physics sigma meson" of the relativistic mean field theories, but with a well defined chiral status.  The second one, named $\sigma'$ as sometimes in Bonn-like OBE approaches, is  a fictitious object simulating two-pion (or two-rho) exchange not reducible to the iteration of the single pion (or single rho) exchange with only nucleons in the intermediate states. 
It is represented by the $\Delta$ box diagrams of fig. \ref{SIGEXCH}b. It turns out that these two objects contribute with similar strength to the $NN$ attraction in the nuclear medium.\\
 A remarkable point is that each of these two "sigma mesons" can be associated with a specific three-body force, which is undoubtedly the main origin of the saturation mechanism. In effect the combination of the response (e.g. the self-consistent modification of the confined quark wave function) of the nucleon to the nuclear scalar field (fig. \ref{THREEBODY}a) and of the cubic $s^3$ tadpole term in the chiral potential (fig. \ref{THREEBODY}b) generates a contribution to the binding energy per nucleon, $E^{(3b-s)}$,  given by Eq. \eqref{eq:THREEBODN}. We recall that this 
 effect is governed by two QCD-connected parameters, $C$ and $C_\chi$. As already mentioned above the diagram of fig. \ref{THREEBODY}c accounts for the Pauli-blocking in the $p-h$ polarization bubble of fig. \ref{SIGEXCH}, which would be lacking if the $\Delta$ box diagrams were replaced by a fictitious $\sigma'$ meson. It can also be seen as a correction of the spin-isospin Fock term due to the dressing of the screened $\pi+\rho$ exchange by the $\Delta-h$
 virtual excitations (see fig. \ref{THREEBODY}c, lower panel). The corresponding contribution to the binding energy can be obtained as the difference: 
 \begin{equation}
 E^{(3b-\sigma')} = E_H^{\Delta\, box} -E_H^{sigma'}.  
 \end{equation}
 We see in fig. \ref{THREEBODYCC} that $E^{(3b-s)}$ and  $E^{(3b-\sigma')}$ are very similar in magnitude and both have a density dependence close to $\rho^2$
 as it should for a pure three-body force.
\section{Summary and conclusion}
The set of works carried out with Peter Schuck in the nineties and dealing with the modification of the pion-pion interaction in the nuclear medium provided the main stimulus  of an approach now known as the chiral confining model. It has progressively emerged  as based on three important physical pillars. First, the chiral invariant scalar field $s$, associated with the radial fluctuation of the chiral condensate, is identified with the "nuclear physics sigma meson" of relativistic theories \cite{Chanfray2001,Martini2006}. Second, the effect of the quark substructure  of the nucleon is reflected by its polarizability in the presence of the nuclear scalar field generating a repulsive three-nucleon force providing an efficient saturation mechanism \cite{Chanfray2005,Chanfray2007,Massot2008,Rahul}. Third, the associated response parameters, namely, the nucleon scalar coupling constant, $g_S$, and scalar susceptibility, $\kappa_{NS}$, can be related to two chiral properties of the nucleon given by lattice QCD simulations, imposing severe constraints on their values \cite{Chanfray2007,Massot2008,Rahul}. It was originally built  using  a linear sigma model  (L$\sigma$M) but subsequently enriched by replacing the  L$\sigma$M chiral potential by its NJL equivalent \cite{Chanfray2011,Chanfray2023}, making lattice QCD constraints more compatible with the chiral properties obtained in confining models of the nucleon. 
A further step  has been recently  accomplished \cite{Universe} using  an effective Hamiltonian inspired by QCD that allows the simultaneous introduction, within some ansatz prescriptions, of a confining model for the nucleon allowing the calculation of the response parameters while generating an equivalent NJL model. 
This  QCD-connected  chiral confining model, with inputs  linked to  genuine QCD quantities (string tension $\sigma$, gluon condensate $\mathcal{G}_2$) and well-established hadronic phenomenology ($\kappa_\rho\sim 6$), is able to generate, via an adapted G matrix approach, a pair correlation function that ensures the natural UV regularization of the pionic  correlation energy. It also gives, without further fine-tuning, reasonable results for the chiral properties of the nucleon ($a_2$ and $a_4$ parameters extracted from lattice data) and saturation properties of nuclear matter, once the effect of short-range correlations (G matrix) is properly implemented. Using this QCD-connected model as un input, we have proposed in this paper an improved version of the contribution of $2\pi$ loops  to the binding energy with a clarification of the real origin of the scalar-isoscalar attraction between nucleon in the nuclear medium. It is  illustrated by the  "two-sigma-mesons" picture, one being related to the chiral invariant  $s=\sigma_W$, and the other one to the $\Delta$ box diagrams. Moreover the associated three-body forces presumably constitute  the dominant saturation mechanism.  Hence this approach provides a link between the chiral properties of the nucleon and the saturation mechanism and/or a link between the fundamental theory of strong interaction and nuclear matter properties, although the results presented in this paper  should be enriched in various aspects of this multi-step approach for application to high density nuclear matter and the neutron star equation of state. Given the remaining theoretical uncertainties concerning both the many-body treatment and the nucleon modeling, the  strategy presently adopted by the Lyon group  is to use a Bayesian method with QCD-connected parameters and lattice data ($a_2, a_4$) parameters, implemented with their associated uncertainties,  as   prior input variables as described in two recent papers \cite{Rahul,Cham}.

\section*{acknowledgments}I acknowledge my long term collaborators, M. Ericson, M. Martini, D. Davesne, H. Hansen and  also more recent ones,  J. Margueron and M. Chamsedinne, who where involved at various stages of the theoretical developments described in this paper.

\bibliographystyle{spphys}       
\bibliography{biblio}

\begin{thebibliography}{999}
\bibitem{Schuck1988} P. Schuck,  W. N\"{o}renberg and G. Chanfray,
Zeit. fur Phys. A330, 119 (1988).
\bibitem{Chanfray1990} G. Chanfray, P. Schuck and W. N\"{o}renberg,    Nucl.Phys. A519, 311c (1990). 
\bibitem{Chanfray1991} G. Chanfray, Z. Aouissat, P. Schuck and W. N\"{o}renberg, Phys. Lett. B256, 325 (1991).
\bibitem{Chanfray1992c} G. Chanfray and P. Schuck,
Nucl. Phys. A545,  271c (1992).
\bibitem{Chanfray1993} G. Chanfray and P. Schuck, 
Nucl. Phys. A555, 329 (1993).
\bibitem{Aouissat1993} Z. Aouissat,  P. Schuck and G. Chanfray,
Mod. Phys. Lett. A8, 1379 (1993).
\bibitem{Aouissat1995} Z. Aouissat, R. Rapp, G. Chanfray, P. Schuck and J. Wambach,
Nucl. Phys. A581,  471 (1995).
\bibitem{Rapp1999} R. Rapp, J.W. Durso, Z. Aouissat, G. Chanfray, O. Krehl, P. Schuck, J. Speth and J. Wambach, 
Phys. Rev. C59,  R1237 (1999).
\bibitem{Aouissat2000} Z. Aouissat, G. Chanfray, P. Schuck and J. Wambach,
Phys. Rev. C61,  12202  (2000).
\bibitem{Chanfray1996} G. Chanfray, R. Rapp and J. Wambach,
Phys. Rev. Lett. 76,  368 (1996).
\bibitem{Rapp1998} R.Rapp, G. Chanfray  and J. Wambach,
Nucl. Phys. A617,  472 (1997). 
\bibitem{Urban} M. Urban, M. Buballa, R. Rapp and J. Wambach,    Nucl.Phys. A673, 357 (2000).
\bibitem{RWH00} R. Rapp and J. Wambach, Adv. Nucl. Phys. 25,  1 (2000); H. van Hees and R. Rapp, Phys. Rev. Lett. 97, 102301 (2006).
\bibitem{NA6006} R. Arnaldi et al. (NA60 collaboration), Phys. Rev. Lett. 96,  162302 (2006);  Eur. Phys. J. C61,  711 (2009). 
\bibitem{Chanfray1998} G. Chanfray, J. Delorme and M. Ericson,
Nucl. Phys. A637,  421 (1998).
\bibitem{Rosaclot}G. Chanfray, J. Delorme, M. Ericson and M. Rosa-Clot, Phys. Lett. B 455, 39 (1999).
\bibitem{OSAKA}G. Chanfray, Nucl. Phys. A685,  328c (2001).
\bibitem{PANIC} G. Chanfray, Nucl. Phys. A721,  76c (2003).
\bibitem{Bonnuti}F. Bonnuti et al., Nucl. Phys. A677,  213 (2000); Phys. Rev. Lett. 77,  603 (1996); Phys. Rev.
C60,  018201 (1999).
\bibitem{Staros} A. Starostin et al., Phys. Rev. Lett. 85,  5539 (2000).
\bibitem{TAPS}J.G. Messchendorp et al., Phys. Rev. Lett. 89,  222302 (2002).
\bibitem{SerotWalecka1986} B.D. Serot and J.D. Walecka, Advances in Nucl. Phys. 16, 1 (1986).
\bibitem{Walecka1997} B.D. Serot and J.D. Walecka, Int. J. Mod. Phys. E16, 15 (1997).
\bibitem{Birse}M.C. Birse, Phys. Rev. C 53, R2048 (1996); M.C. Birse and J.A. McGovern, Phys.Lett B309, 231 (1993).
\bibitem{Chanfray2001} G. Chanfray, M. Ericson and P.A.M. Guichon, Phys. Rev. C63, 055202 (2001).
\bibitem{Martini2006} G. Chanfray, D. Davesne, M. Ericson and M. Martini, Eur. Phys. J. A27, 191 (2006).
\bibitem{Boguta83} J. Boguta, Phys. Lett. B120, 34 (1983).
\bibitem{KM74} A.K. Kerman and L.D. Miller in ``Second High Energy Heavy Ion Summer Study'', LBL-3675 (1974). 
\bibitem{BT01} W. Bentz and A.W. Thomas, Nucl. Phys. A696, 138 (2001).
\bibitem{LTY03}D. B. Leinweber, A. W. Thomas and R. D. Young, arXiv:hep-lat/0302020.
\bibitem{LTY04}D. B. Leinweber, A. W. Thomas and R. D. Young, Phys.
Rev. Lett. 92, 242002 (2004).
\bibitem{TGLY04} A. W. Thomas, P. A. M. Guichon, D. B.  Leinweber and R. D. Young, Progr. Theor. Phys. Suppl. 156, 124 (2004); nucl-th/0411014.
\bibitem{AALTY10} W. Armour, C.R. Alton, D. B. Leinweber, A. W. Thomas and R. D. Young, Nucl. Phys. A840, 97 (2010).
\bibitem{Ericson2007}M. Ericson and G. Chanfray, Eur. Phys. J. A 34, 215 (2007).
\bibitem{Guichon1988} P.A.M. Guichon, Phys. Lett. B200, 235 (1988).
\bibitem{Guichon2004} P.A.M. Guichon and A.W. Thomas, Phys. Rev. Lett. 93, 132502 (2004); P.A.M. Guichon, H.H. Matevosyan, N. Sandulescu and A.W. Thomas, Nucl. Phys. A772, 1 (2006); R. Stone, P.A. M. Guichon, P. G. Reinhard and A.W. Thomas,  Phys. Rev. Lett. 116, 092511 (2016).
\bibitem{Chanfray2005} G. Chanfray and M. Ericson, Eur. Phys. J. A 25, 151 (2005).
\bibitem{Chanfray2007}G. Chanfray and M. Ericson, Phys. Rev. C75  015206 (2007).
\bibitem{Massot2008} E. Massot and G. Chanfray, Phys. Rev. C78, 015204 (2008).	
\bibitem{Massot2009} E. Massot and G. Chanfray, Phys. Rev. C80, 015202 (2009).
\bibitem{Rahul} R. Somasundaram, J. Margueron, G. Chanfray and H. Hansen, Eur. Phys. J. A 58, 5 (2022).
\bibitem{Massot2012} E. Massot, J. Margueron and G. Chanfray, EPL 97, 39002 (2012).
\bibitem{Chanfray2011} G. Chanfray and M. Ericson, Phys. Rev. C75, 015204 (2011).
\bibitem{Cham} M. Chamseddine, J. Margueron, G. Chanfray, H. Hansen and  R. Somasundaram, to be published in Eur. Phys. J. A; arXiv:2304.01817 [nucl-th].
\bibitem{Chanfray2023} G. Chanfray, H. Hansen and J. Margueron, arXiv:2304.01036 [nucl-th].
\bibitem{Universe} G. Chanfray, M. Ericson and M. Martini,     Universe 9 (2023) 7, 316; arXiv : 2307.03484 [nucl-th]
\bibitem{Simonov1997} Yu. A. Simonov, Phys. At. Nucl. 60, 2069 (1997; Few-BodySyst. 25, 45 (1998).
\bibitem{Tjon2000} Yu. A. Simonov and J. A. Tjon, Phys. Rev. D62, 014501 (2000).
\bibitem{Simonov2002} Yu. A. Simonov, J. A. Tjon and J. Weda, Phys. Rev. D65, 094013 (2002). 
\bibitem{Simonov2002a} Yu. A. Simonov, Phys. Rev. D65, 094018 (2002).
\bibitem{Simonov-light} Yu. A. Simonov, Phys. At.Nucl. 67, 846 (2004), hep-ph/0302090;
Yu. A. Simonov, Phys. At. Nucl. 67, 1027 (2004), hep-ph/0305281;
Yu. A. Simonov, Int. Mod. Phys. A 31, 165016 (2016), arXiv:1509.06930 [hep-ph].
\bibitem{Digiacomo} A. Di Giacomo, H.G. Dosch, V.I. Shevchenko and Yu. A. Simonov, Physics Reports 372,  319 (2002).
\bibitem{BBRV98} P. Bicudo, N. Brambilla, E. Ribeiro and A. Veiro, Phys. Lett.  B442, 349 (1998).
\bibitem{CSN1990} G. Chanfray, P. Schuck and W. N\"{o}renberg,  Nucl. Phys. A519,  311c (1990).
\bibitem{HP75} G. Hohler and E. Pietarinen, Nucl. Phys. B95, 210 (1975).
\bibitem{MHE89} Machleidt, K. Holinde and Ch. Elster, Physics Reports 149, 1 (1987).
\bibitem{BONNPOT} R. Brockmann and R. Machleidt, Phys. Rev. C42, 1965 (1990).
\bibitem{Ichimura} M. Ichimura, H. Sakai, and T. Wakasa, Prog. Part. Nucl. Phys. 56, 446 (2006).
\bibitem{Friman} B. L. Friman and E. M. Nyman, Nucl. Phys. A302, 365 (1978).
\bibitem{Goldstone} J. Goldstone, Proc. Roy. Soc. A239,  267 (1956).
\bibitem{CEM2022} G. Chanfray, M. Ericson and M. Martini, Eur. Phys. J. ST 230, 4357 (2021).
\bibitem{Schuck89} P. Schuck et al, Progr. Part. Nucl. Phys. 22, 181 (1989).
\bibitem{Dewulf} Y. Dewulf, W. H. Dickhoff, D. Van Neck, E. R. Stoddard, and M.Waroquier, Phys. Rev; . Lett. 90,  152501 (2003).
\bibitem{Ciofi} C. Ciofi degli Atti and S. Liuti, Phys. Lett. B225,  215 (1989);
C. Ciofi degli Atti and S. Liuti, Phys. Rev. C41, 1100 (1990); C. Ciofi degli Atti,  Nucl. Phys. A532, 241 (1991).
\bibitem{EMC84} R.G. Arnold et al., Phys. Rev. Lett. 52, 727  (1984);
 J. Ashman et al., European Muon Collaboration, Phys. Lett. B202, 1004, (1988);
 S.V. Akulinichev, S.A. Kulagin and G.M. Vagradov, Phys. Lett. B158,  485 (1985).

\bibitem{Holinde} K. Holinde, Phys. Rep. 68, 121 (1981). 
\bibitem{Baldo2012} M. Baldo, H.R. Moshfegh, Phys. Rev. C86, 024306 (2012).
\bibitem{Chanfray1985} G. Chanfray, J. Delorme and M. Ericson, Phys. Rev. C31, 1582 (1985).
\bibitem{TB1} J.-I. Fujita and H. Miyazawa, Prog. Theor. Phys. 17, 360 (1957).
\bibitem{TB2} G.  Brown and A.M Green, Nucl. Phys. A137 (1969). 
\bibitem{TB3} S. A. Coon, M. D. Scadron, P. C. McNamee, B. R. Barrett, D. W. E. Blatt. and B. H. J. McKeller, Nucl. Phys. A317, 242 (1979).
\bibitem{TB4} H. T. Coelho, T. K. Das, and M. R. Robilotta, Phys. Rev. C 28, 1812 (1983); M. R. Robilotta and H. T. Coelho, Nucl.
Phys. A460, 645 (1986).
\bibitem{CEFT} R. Machleidt and D. R. Entem,   Phys.Rep. 503, 1 (2011); arXiv:1105.2919 [nucl-th].
\bibitem{KFW02} N.  Kaiser, S.  Fritsch and W.  Weise,  Nucl. Phys. A697, 82 (2002).

\end{thebibliography}

\end{document}